\documentclass[twocolumn,english,aps,pra,tightenlines,longbibliography]{revtex4-1}
\usepackage{lmodern}
\usepackage{color}
\usepackage{textcomp}
\usepackage{mathtools}
\usepackage{amsmath}
\usepackage{amssymb}
\usepackage{graphicx}

\makeatletter
\usepackage{microtype}

\usepackage[unicode=true,pdfusetitle,
 bookmarks=true,bookmarksnumbered=false,bookmarksopen=false,
 citecolor=cyan,urlcolor=magenta,linkcolor=blue, breaklinks=false,
 pdfborder={0 0 0},backref=false,colorlinks=true]{hyperref}

\usepackage{amsthm}\usepackage{epstopdf}\usepackage{color}
\usepackage{bm}\usepackage{babel}
\usepackage{upgreek}
\usepackage{textcomp}
\usepackage{siunitx}
\newcommand\figref[2]{Fig.\,\ref{#1}\hyperref[#1]{(#2)}}

\newcommand\fref[2]{\ref{#1}\hyperref[#1]{(#2)}}

\makeatother

\usepackage{babel}

\begin{document}
\title{Magnetic-free traveling-wave nonreciprocal superconducting microwave
components}
\author{Dengke Zhang$^{1,2,}$}
\author{Jaw-Shen Tsai$^{1,2,}$}
\email{tsai@riken.jp}

\affiliation{$^{1}$Department of Physics, Tokyo University of Science, Kagurazaka, Tokyo 162-8601, Japan \\
$^{2}$RIKEN Center for Quantum Computing, RIKEN, Wako, Saitama 351-0198, Japan}
\begin{abstract}
We propose a design to realize integrated broadband nonreciprocal
microwave isolators and circulators using superconducting circuit
elements without any magnetic materials. To obtain a broadband response,
we develop a waveguide-based design by temporal modulations. The corresponding
compact traveling-wave structure is implemented with integrated superconducting
composite right-/left-handed transmission lines. The calculations
show that the bandwidth of \SI{580}{\MHz} can be realized over a
nonreciprocal isolation of \SI{20}{\dB} in reflections. Such on-chip
isolators and circulators are useful for cryogenic integrated microwave
connections and measurements, such as protecting qubits from the amplified
reflected signal in multiplexed readout.
\end{abstract}
\maketitle

\section{Introduction}

Realizing on-chip nonreciprocal isolation is a significant difficulty
in microwave integrated circuits \citep{Nagulu2020NE,Kord2020PI,Jalas2013NP}.
The need to overcome this difficulty has recently become increasingly
urgent for on-chip microwave quantum information processing, especially
for low-noise quantum measurements \citep{Ranzani2019IMM,Lupascu2007NP,Mallet2009NP,Jeffrey2014PRL,Siddiqi2004PRL,Yamamoto2008APL,OBrien2014PRL,Macklin2015S,Kamal2011NP,Abdo2013PRX,Metelmann2015PRX,Abdo2014PRL,Lau2018NC,Lecocq2017PRA,Abdo2019NC}.
Achieving nonreciprocity requires breaking time-reversal symmetry,
which can be accomplished via the magneto-optical effect, nonlinearity,
or spatiotemporal modulation \citep{Nagulu2020NE,Kord2020PI,Yu2009NP,Kamal2011NP,Viola2014PRX}.
Although each one has its own advantages, on-chip integration remains
a huge challenge because the reported implementations so far suffer
from high complexity, intricate control, small bandwidth, or large
footprint. Achieving the magneto-optical effect requires magneto-optical
materials, such as ferrites, which are difficult to be integrated
by the standard process and unavoidably introduce unwanted magnetic
noise \citep{Shoji2016JO,Wang2019PRLb}. Nonreciprocity using nonlinearity
is power-dependent, which means that the nonreciprocal performance
depends on input power; furthermore, some reciprocal leakages happen
with multi-inputs \citep{Shi2015NP,Bino2018O,RosarioHamann2018PRL}.
In recent years, the method of spatiotemporal modulation has shown
powerful vitality because diverse modulations can be provided by electric,
acoustic, or mechanical dynamic biasing \citep{Nagulu2020NE,Kord2020PI,Sounas2017NP,Kamal2011NP,Reiskarimian2016NC,Rosenthal2017PRL,Chamanara2017PRB,Yu2019JMS,Toth2017NP,Shen2018NC,Barzanjeh2017NC,Peterson2017PRX}. 

In general, the spatiotemporal modulation requires simultaneously
spatially and temporally dynamic modulations of the material parameters,
such as the refractive index \citep{Yu2009NP,Lira2012PRL,Sounas2017NP,Chamanara2017PRB}.
In resonance-based configurations, such modulations are implemented
by applying spatially uniform modulation but specific timing control
on each resonator or coupling segment \citep{Kamal2011NP,Abdo2013PRX,Sliwa2015PRX,Estep2014NP,Chapman2017PRX}.
The complexities of these types of structure and controls are partly
eased at the expense of working bandwidth. To obtain a broadband nonreciprocal
response, waveguide-based implementations are needed. Common spatiotemporal
modulation for a waveguide is very difficult for fabricating the structure
or applying the modulation \citep{Yu2009NP,Lira2012PRL,Choi2017NC}.
Specifically, such a spatiotemporal modulation can also be achieved
by a strong microwave field for a microwave isolator or circulator.
Ranzani \textit{et al. }have experimentally demonstrated a broadband
isolator using a Josephson-junction transmission line \citep{Ranzani2017PRA}.
In their scheme, an intraband transition was achieved by spatiotemporal
modulation with a microwave pumping. As the frequency of pumping (or
modulation) is very high (comparable to the frequency of signal),
it is not easy to well terminate both modes simultaneously, the frequencies
of which are spaced by a value of the modulation frequency. 

Recently, a proposal has suggested that a broadband nonreciprocal
isolation can be realized by applying two spatially uniform modulations
on two waveguide-based mode transformers, in which the interband transition
between two modes is required by means of a dynamic modulation \citep{Fang2012PRL,Tzuang2014NP}.
With such an approach, the requirement of spatiotemporal modulation
is reduced to two spatially discrete temporal-modulations. Thus, the
research priority becomes how to elegantly realize a temporally modulated
mode transformer with waveguides. Moreover, the termination impedance
matching for both modes are required to eliminate the reflections.
In a superconducting circuit, the couplings between two common coplanar
waveguides (CPWs) are very weak, which makes it difficult to achieve
the required mode transformation within a small footprint. In addition,
the interband transition cannot be realized by a simple modulation
of coupled CPWs. Here, we propose a new design of integrated microwave
isolator/circulator with superconducting composite right-/left-handed
(CRLH) transmission lines (TLs) \citep{Lai2004IMM,Hirota2009ITMTT}.
There are several merits to using CRLH TLs: First, the supported light
possesses a small effective wavelength, which means that a strong
coupling can be realized between such two TLs and, meanwhile, a small
footprint can also be achieved. Second, the coupled CRLH TLs can be
modulated in a low frequency, which implies the frequency gap of coupled
two modes is small. This feature will be a benefit to realizing a
successful termination impedance matching for both modes. Third, CRLH
TLs are one kind of metamaterials-based waveguides that holds a high
design degree of freedom. 

In this work, we design a structure for realizing the broadband nonreciprocal
isolation with the CRLH TLs. In the designed structure, the required
mode transformation is accomplished with two coupled CRLH TLs. The
entire designed structure can be fabricated with a standard superconducting
circuit process, and the spatially uniform modulations are readily
applied using electric local biasing lines. This paper is organized
as follows. In Section \ref{secII}, we present the structure design
and the corresponding model, and an intuitive operation principle
is given with the help of Ramsey interference. In Section \ref{secIII},
we introduce the CRLH TLs and their couplings. Then a 3-dB CRLH directional
coupler is implemented and modulated, as described in Section \ref{secIV}.
Section \ref{secV} shows the nonreciprocal responses of the entire
designed CRLH-based structure and Section \ref{secVI} provides a
possible design layout with superconducting lines and Josephson junctions,
and some discussions about the improvement. Last, we present the conclusion
in Section \ref{secVII}.

\section{Structure design and model\label{secII}}

\begin{figure}[ht]
\centering \includegraphics{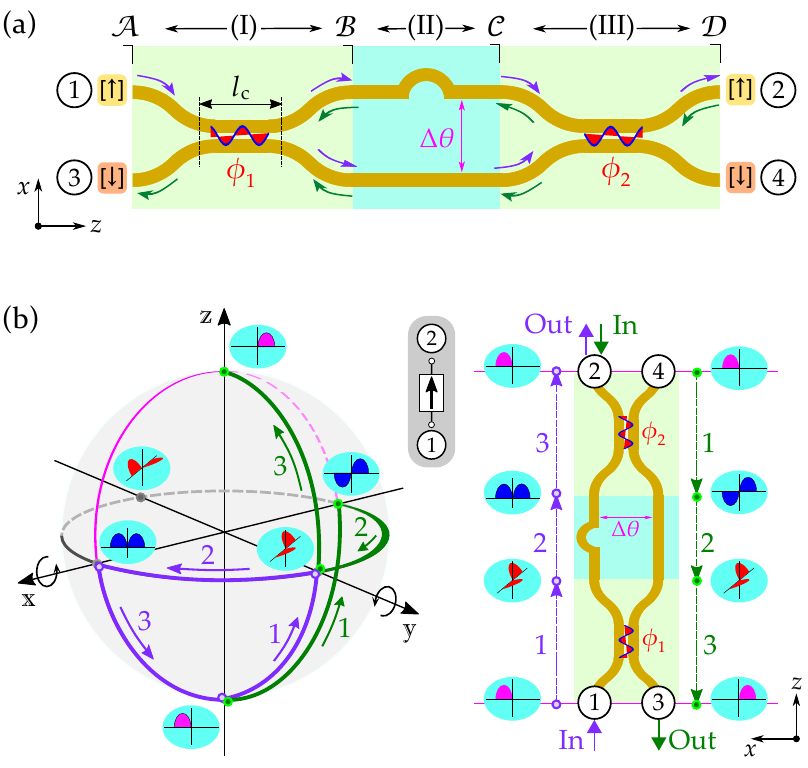} \caption{Schematic of the design and working principle. (a) The designed structure
is constructed with a cascade of three stages of dual-waveguide configuration,
where the two bare waveguides support guided mode-$[\uparrow]$ and
mode-$[\downarrow]$, respectively. Stage-I and stage-III are 3-dB
forward-wave directional couplers, the coupling coefficients of which
are temporally modulated with initial phases of $\phi_{1}$ and $\phi_{2}$,
respectively. Stage-II is two uncoupled waveguides with different
propagation lengths to introduce a phase difference of $\Delta\theta$.
The entire structure performs as a four-port network, which is indexed
by port-1 to port-4. (b) Field evolutions are displayed on a Bloch
sphere (left) and in real space (right) with different injections
at port-1 and port-2. Here, we set $\{\phi_{1},\Delta\theta,\phi_{2}\}=\{0,\pi/2,\pi/2\}$.
The light injected at port-1 will be output via port-2, whereas the
input light at port-2 will go to port-3. These behaviors indicate
that a nonreciprocal isolation between port-1 and port-2 can be achieved
with the design shown in (a). \label{fig1}}
\end{figure}

The proposed structure contains three stages: a 3-dB forward-wave
directional coupler (stage-I: plane-$\mathcal{A}$ to plane-$\mathcal{B}$),
then a phase difference between the two split beams introduced through
two uncoupled waveguides (stage-II: plane-$\mathcal{B}$ to plane-$\mathcal{C}$),
and finally followed by another 3-dB forward-wave directional coupler
(stage-III: plane-$\mathcal{C}$ to plane-$\mathcal{D}$), as shown
in \figref{fig1}{a}. Importantly, the two 3-dB forward-wave directional
couplers have a time-varying coupling coefficient via dynamic modulation.
As shown in \figref{fig1}{a}, we consider that each coupler consists
of two waveguides, whose eigenmodes are mode-$[\uparrow]$ and mode-$[\downarrow]$
propagating along the $z$-direction, respectively. Assume that the
time-varying coupling coefficient between mode-$[\uparrow]$ and mode-$[\downarrow]$
is expressed by

\begin{equation}
K(t)=K_{0}\cos(\Omega_{\mathrm{m}}t+\phi),\label{eqModu.coupling}
\end{equation}
where $K_{0}$ is the maximum coupling coefficient (i.e., the amplitude
of $K$), $\Omega_{\mathrm{m}}$ is the modulation frequency, and
$\phi$ is the modulation initial phase. In this work, the 3-dB beam
coupling means that there is a relation of $|K_{0}l_{\mathrm{c}}|=\pi/4$
with coupling length $l_{\mathrm{c}}$ of the coupler. In the frame
rotating with the uncoupled waveguides' propagation constants, the
equation of motion of fields in the directional coupler can be written
with rotating wave approximation (see Appendix A for the details)
as

\begin{equation}
{\displaystyle \mathrm{j}\frac{\mathrm{\partial}}{\partial z}\left[\begin{array}{c}
a_{\uparrow}\\
a_{\downarrow}
\end{array}\right]=\left[\begin{array}{cc}
0 & K_{0}\mathrm{e}^{-\mathrm{j}\phi}\\
K_{0}\mathrm{e}^{\mathrm{j}\phi} & 0
\end{array}\right]\left[\begin{array}{c}
a_{\uparrow}\\
a_{\downarrow}
\end{array}\right]},\label{eqHam.coupler}
\end{equation}
where $a_{\uparrow}$ and $a_{\downarrow}$ represent field amplitudes
of mode-$[\uparrow]$ and mode-$[\downarrow]$, respectively \citep{Fang2012PRL,Qin2018PRL}.
By solving Eq.\,(\ref{eqHam.coupler}), we can deduce the spatial
evolutions of mode-$[\uparrow]$ and mode-$[\downarrow]$ along the
$z$-direction

\begin{equation}
\left[\begin{array}{c}
a_{\uparrow}(z)\\
a_{\downarrow}(z)
\end{array}\right]=\left[\begin{array}{cc}
{\displaystyle \cos\left(K_{0}z\right)} & -\mathrm{j}\mathrm{e}^{-\mathrm{j}\phi}{\displaystyle \sin\left(K_{0}z\right)}\\
-\mathrm{j}\mathrm{e}^{\mathrm{j}\phi}{\displaystyle \sin\left(K_{0}z\right)} & {\displaystyle \cos\left(K_{0}z\right)}
\end{array}\right]\left[\begin{array}{c}
a_{\uparrow}(0)\\
a_{\downarrow}(0)
\end{array}\right].\label{eqSpace.evolution}
\end{equation}
According to the definition of the scattering matrix of a two-port
network, the scattering matrix {[}$\mathbf{M}^{(\mathrm{c})}${]}
of the directional coupler can be obtained straightforwardly through
Eq.\,(\ref{eqSpace.evolution}). For the 3-dB coupler in stage-I,
using $K_{0}l_{\mathrm{c}}=\pi/4$, we can express the scattering
matrix (relation of fields between ports in plane-$\mathcal{A}$ and
ports in plane-$\mathcal{B}$) as

\begin{equation}
\mathbf{M}^{(\mathrm{c})}(\phi_{1})=\frac{1}{\sqrt{2}}\left[\begin{array}{cc}
{\displaystyle 1} & -\mathrm{j}\mathrm{e}^{-\mathrm{j}\phi_{1}}\\
-\mathrm{j}\mathrm{e}^{\mathrm{j}\phi_{1}} & 1
\end{array}\right],\label{eqScatM.coupler}
\end{equation}
where $\phi_{1}$ is the modulation initial phase of the 3-dB coupler
in stage-I. In comparison with the scattering matrix of a common (unmodulated)
3-dB directional coupler, we can find that a phase shift $\phi_{1}$
is appended to the cross-coupling parameter of the modulated coupler
(i.e., transition between mode-$[\uparrow]$ and mode-$[\downarrow]$),
resulting from the dynamic modulation of the coupling coefficient.
Regarding the coupler in stage-III (from plane-$\mathcal{C}$ to plane-$\mathcal{D}$),
the corresponding scattering matrix $\mathbf{M}^{(\mathrm{c})}(\phi_{2})$
can be readily given with the replacement of $\phi_{1}$ by $\phi_{2}$
in Eq.\,(\ref{eqScatM.coupler}). For the two uncoupled waveguides
in stage-II, the scattering matrix can be represented by

\begin{equation}
\mathbf{M}^{(\mathrm{b})}(\Delta\theta)=\left[\begin{array}{cc}
\mathrm{e}^{\mathrm{j}\Delta\theta} & 0\\
0 & 1
\end{array}\right],\label{eqSactM.Phaseshift}
\end{equation}
where $\Delta\theta$ is the introduced phase difference between propagating
waves of mode-$[\uparrow]$ and mode-$[\downarrow]$.

By using a cascade of the scattering matrices of the three stages,
we can easily obtain the scattering matrix of the entire structure.
For the forward ($+z$-direction) propagation waves, the scattering
matrix is written as

\begin{equation}
\mathbf{M}^{(\Rightarrow)}=\mathbf{M}^{(\mathrm{c})}(\phi_{2})\mathbf{M}^{(\mathrm{b})}(\Delta\theta)\mathbf{M}^{(\mathrm{c})}(\phi_{1}),\label{eqScatM.Forward}
\end{equation}
and for the backward ($-z$-direction) propagation waves, the corresponding
scattering matrix is expressed as 

\begin{equation}
\mathbf{M}^{(\Leftarrow)}=\mathbf{M}^{(\mathrm{c})}(\phi_{1})\mathbf{M}^{(\mathrm{b})}(\Delta\theta)\mathbf{M}^{(\mathrm{c})}(\phi_{2}).\label{eqScatM.Backward}
\end{equation}
Here, we assume that $\{\phi_{1},\Delta\theta,\phi_{2}\}=\{0,\pi/2,\pi/2\}$;
the two specific scattering matrices are calculated as

\begin{equation}
\mathbf{M}^{(\Rightarrow)}=\left[\begin{array}{cc}
\mathrm{j} & 0\\
0 & 1
\end{array}\right],\quad\mathbf{M}^{(\Leftarrow)}=\left[\begin{array}{cc}
0 & -\mathrm{j}\\
1 & 0
\end{array}\right].\label{eqNonreciprocal}
\end{equation}
From Eq.\,(\ref{eqNonreciprocal}), we can find that a forward injected
field of mode-$[\uparrow]$ can be maintained finally, whereas a backward
injected field of mode-$[\uparrow]$ is transferred to the field of
mode-$[\downarrow]$, which are shown by paths with purple and green
arrows in \figref{fig1}{a}, respectively. That is, the transmission
of light from port-1 to port-2 is allowed, but the reverse process
is suppressed (the input light at port-1 goes to port-3). Such a behavior
implies that a nonreciprocal isolation can be achieved using our design
with the specific settings for $\{\phi_{1},\Delta\theta,\phi_{2}\}$. 

As a dual-mode system, a geometric description will be useful for
understanding the working principle \citep{Ranzani2014NJP,Yurke1986PRA,Leonhardt2010Book}.
For our structure, the processes of mode transformations can be described
by a photonic Ramsey interference. The function of a 3-dB directional
coupler is to apply a ``$\pi/2$-pulse'' excitation to the input
field of mode-$[\uparrow]$ or mode-$[\downarrow]$. After passing
through the 3-dB directional coupler, the input field will be transferred
to a superposition field of mode-$[\uparrow]$ and mode-$[\downarrow]$.
This means that the mode transformation is achieved by the 3-dB directional
coupler. Stage-II works like a free evolution of the superposition
state after a detuned $\pi/2$-pulse driving, which is realized by
introducing a phase difference between field components of mode-$[\uparrow]$
and mode-$[\downarrow]$. Finally, another $\pi/2$-pulse is applied
by the 3-dB directional coupler in stage-III. In particular, the modulation
initial phases ($\phi_{1}$ and $\phi_{2}$) applied in the couplers
play the role of the initial phase of a $\pi/2$-pulse signal. For
the special setting of $\{\phi_{1},\Delta\theta,\phi_{2}\}=\{0,\pi/2,\pi/2\}$,
the field evolutions on a Bloch (or Poincar\'{e}) sphere and in real
space are displayed in \figref{fig1}{b}. The forward evolution
with the input field at port-1 is represented by purple curves, and
the backward evolution with the input field at port-2 is denoted by
green curves. These demonstrations clearly show that the isolation
response between port-1 and port-2 is realized by performing such
a three-stage spatial operation.

To implement such a structure in a superconducting circuit, a real
3-dB directional coupler is needed; moreover, the coupling coefficient
can be modulated. However, it is unrealistic to achieve an on-chip
3-dB directional coupler using common CPWs, because the coupling between
two parallel superconducting CPWs is very weak. To realize such a
compact 3-dB directional coupler on a chip, CRLH TLs are introduced
in this work.

\section{CRLH transmission lines\label{secIII}}

\begin{figure}[ht]
\centering \includegraphics{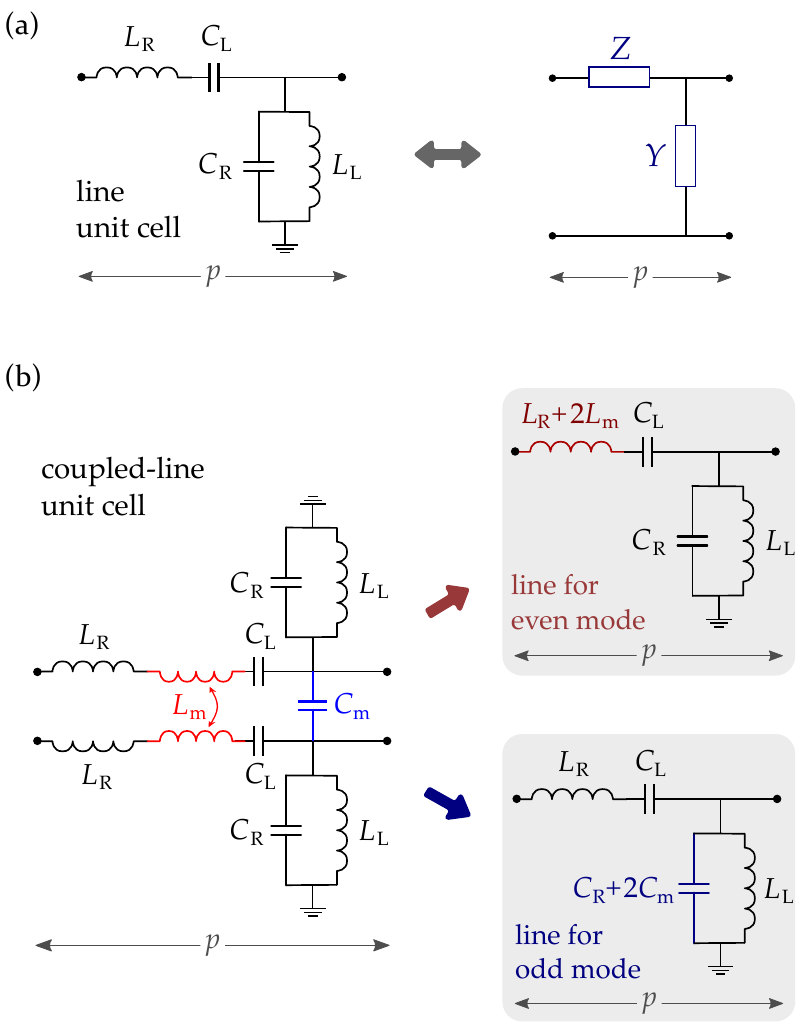} \caption{Unit cells of CRLH TL and coupled-line. (a) Circuit model of a unit
cell of CRLH TL. (b) Circuit model of unit cells of the CRLH directional
coupler and the corresponding even- and odd-mode TL models. The even-
and odd-mode equivalent circuits have the same topology but with different
substitutions $L_{\mathrm{R}}\rightarrow L_{\mathrm{R}}+2L_{\mathrm{m}}$
and $C_{\mathrm{R}}\rightarrow C_{\mathrm{R}}+2C_{\mathrm{m}}$, respectively.
\label{fig2}}
\end{figure}

\subsection{Periodic CRLH TLs}

Here, we use the CRLH TLs to realize a compact broadband forward-wave
directional coupler. The reason is that there are two benefits with
CRLH TLs: first, the wavenumber of guided modes in CRLH TLs can be
very large, which means that the effective wavelength is very small;
second, the strong coupling between such two CRLH TLs can be realized
by introducing optimal mutual inductance or capacitance \citep{Lai2004IMM}.
In reality, an equivalent CRLH TL is constructed by periodically repeating
a small lumped unit cell with pitch $p$, as shown in \figref{fig2}{a}.
The unit cell consists of the combination of right-handed elements
of series inductance $L_{\mathrm{R}}$ and shunt capacitance $C_{\mathrm{R}}$with
left-handed elements of shunt inductance $L_{\mathrm{L}}$ and series
capacitance $C_{\mathrm{L}}$ (see Refs. \citep{Lai2004IMM,Mmongia2007Book,Martin2015Book}).
For simplicity, considering the lossless TLs, the propagation constant
$\beta$ satisfies the relation $\beta=-\mathrm{j}\gamma$, where
$\gamma$ is the complex propagation constant. According to the Bloch-Floquet
theorem, the dispersion relation of a CRLH TL can be obtained by applying
periodic boundary conditions to a unit cell in the form

\begin{equation}
\beta(\omega)=\frac{1}{p}\cos^{-1}\left[1+\frac{Z(\omega)Y(\omega)}{2}\right],\label{eqWavenumber}
\end{equation}
where the series impedance ($Z$) and shunt admittance ($Y$) of the
unit cell are respectively given by 
\begin{align}
Z(\omega) & =\mathrm{j}\left(\omega L_{\mathrm{R}}-\frac{1}{\omega C_{\mathrm{L}}}\right),\label{eqSeriesImpe}\\
Y(\omega) & =\mathrm{j}\left(\omega C_{\mathrm{R}}\text{\textminus}\frac{1}{\omega L\mathrm{_{L}}}\right).\label{eqShuntAdmi}
\end{align}
Furthermore, the characteristic impedance ($Z_{0}$) of a CRLH TL
is read as 
\begin{equation}
Z_{0}(\omega)=\sqrt{Z(\omega)/Y(\omega)}.\label{eqCharImpe}
\end{equation}
As a comparison demonstrated in Appendix B, a homogeneous TL with
$\beta,Z_{0}$ can effectively model the periodic CRLH TL in the long-wavelength
limit.

\begin{figure}
\centering \includegraphics[width=8.5cm]{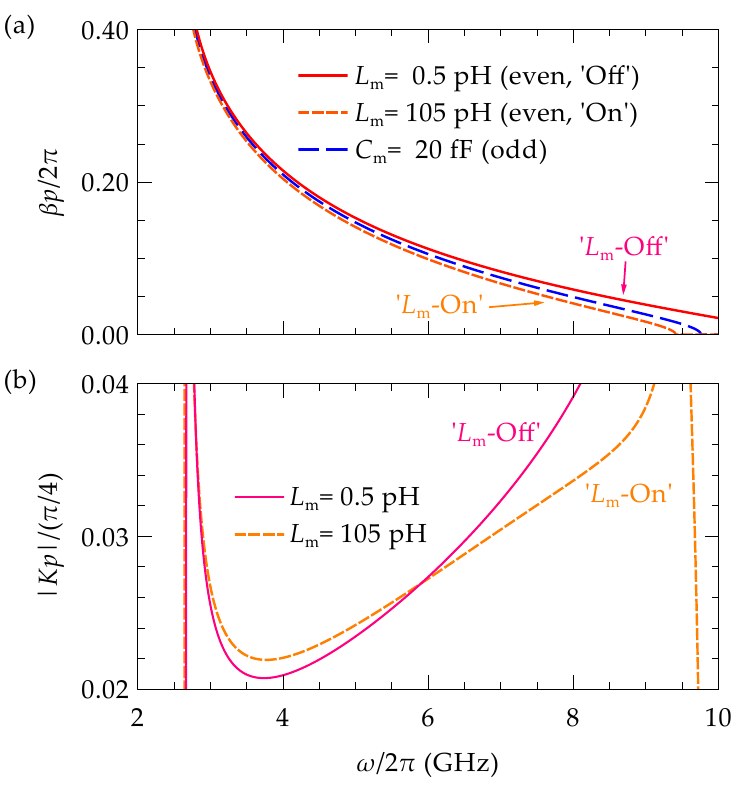} \caption{Dispersion relations and coupling coefficients. (a) Dispersion curves
for odd-mode with $C_{\mathrm{m}}=\SI{20}{\fF}$ and even-mode with
$L_{\mathrm{m}}=\SI{0.5}{\pico\henry}$ and $\SI{105}{\pico\henry}$.
The lower $L_{\mathrm{m}}$ is called '$L_{\mathrm{m}}$-Off' and
the upper $L_{\mathrm{m}}$ is called '$L_{\mathrm{m}}$-On'. (b)
Frequency dependence of magnitudes of coupling coefficients under
the conditions of '$L_{\mathrm{m}}$-Off' and '$L_{\mathrm{m}}$-On',
in which the odd-mode is fixed with $C_{\mathrm{m}}=\SI{20}{\fF}$
in the calculations. \label{fig3}}
\end{figure}

\subsection{Coupling of two identical CRLH TLs}

A CRLH directional coupler can be constructed by introducing mutual
inductance $L_{\mathrm{m}}$ or capacitance $C_{\mathrm{m}}$ into
two CRLH TLs, as shown in \figref{fig2}{b}. Here, we consider a
symmetric directional coupler, where the two bare CRLH TLs are identical.
In a coupled-line, the two bare guided modes are mixed and generate
two hybrid guided modes-- even-mode and odd-mode. The two normal
modes can be analytically reduced to guided modes of two different
bare CRLH TLs. As displayed in \figref{fig2}{b}, the two effective
CRLH TLs for even and odd modes have the same topology as that of
the initial bare CRLH TLs \citep{Caloz2004ITMTT,Hirota2009ITMTT}.
The difference between the two effective CRLH TLs is $L_{\mathrm{R}}$
revised by the addition of $2L_{\mathrm{m}}$ for the even-mode and
$C_{\mathrm{R}}$ revised by the addition of $2C_{\mathrm{m}}$ for
the odd-mode. As the topologies of even/odd lines are the same as
that of a bare CRLH TL, the analyses of the CRLH TL can be applied
directly to calculate the corresponding parameters of even and odd
modes with appropriate substitutions in Eqs.\,(\ref{eqWavenumber}--\ref{eqCharImpe}).
We use the notation $\beta_{\mathrm{even}}$, $Z_{\mathrm{even}}$,
$Y_{\mathrm{even}}$, and $Z_{\mathrm{0even}}$ for the even-line
and $\beta_{\mathrm{odd}}$, $Z_{\mathrm{odd}}$, $Y_{\mathrm{odd}}$,
and $Z_{\mathrm{0odd}}$ for the odd-line. According to the coupled-mode
theory, the coupling coefficient of the coupler can be given by (see
Appendix A for the details)

\begin{equation}
K=\frac{\beta_{\mathrm{even}}-\beta_{\mathrm{odd}}}{2}.\label{eqCoupRate.CRLHcoupler}
\end{equation}
The equation tells us that the coupling coefficient can be modulated
if either or both $\beta_{\mathrm{even}}$ and $\beta_{\mathrm{odd}}$
could be changed in a controllable manner. In this work, we will tune
$\beta_{\mathrm{even}}$ by varying the mutual inductance $L_{\mathrm{m}}$
to achieve the modulation of the coupling coefficient $K$. For the
varied $L_{\mathrm{m}}$, we call the lower bound '$L_{\mathrm{m}}$-Off'
and the upper bound '$L_{\mathrm{m}}$-On.' According to the modulation
form of $K$ in (\ref{eqModu.coupling}), the two bound limits will
exactly induce the extreme of $K$, that is, $K_{0}$ and $-K_{0}$,
respectively. This also implies that both the $L_{\mathrm{m}}$-Off
and $L_{\mathrm{m}}$-On will generate the 3-dB coupling if $|K_{0}l_{\mathrm{c}}|=\pi/4$.

In this work, we demonstrate microwave responses around \SI{6}{\GHz},
which is a typical frequency range for the superconducting circuits.
Here, we set $\{L_{\mathrm{R}},C_{\mathrm{R}}\}=\{\SI{300}{\pico\henry},\SI{150}{\fF}\}$
and $\{L_{\mathrm{L}},C_{\mathrm{L}}\}=\{\SI{1400}{\pico\henry},\SI{560}{\fF}\}$
for the bare CRLH TLs. Furthermore, to realize a 3-dB coupler with
an operating frequency of \SI{6}{\GHz}, $C_{\mathrm{m}}$ is fixed
as \SI{20}{\fF}, and meanwhile $L_{\mathrm{m}}$ can take the value
of \SI{0.5}{\pico\henry} or \SI{105}{\pico\henry}, which correspond
to $L_{\mathrm{m}}$-Off and $L_{\mathrm{m}}$-On, respectively. By
using Eq.\,(\ref{eqWavenumber}), we can calculate the dispersion
curve of the odd-mode with $C_{\mathrm{m}}=\SI{20}{\fF}$. Correspondingly,
the dispersion curves of the even-mode with $L_{\mathrm{m}}$-Off
and $L_{\mathrm{m}}$-On are calculated using $L_{\mathrm{m}}=\SI{0.5}{\pico\henry}$
and $\SI{105}{\pico\henry}$, respectively. These results are shown
in \figref{fig3}{a} and indicate that the dispersion curve of the
odd-mode is in the middle of that of the even-mode of $L_{\mathrm{m}}$-Off
and $L_{\mathrm{m}}$-On. By using the expression of $K$ in Eq.\,(\ref{eqCoupRate.CRLHcoupler}),
the magnitude of the coupling coefficient can be changed from maximum
negative to maximum positive when the $L_{\mathrm{m}}$ is continuously
tuned from $L_{\mathrm{m}}$-Off to $L_{\mathrm{m}}$-On. This means
that modulation of the coupling coefficient as the form of Eq.\,(\ref{eqModu.coupling})
is possible if we could temporally vary the $L_{\mathrm{m}}$ between
$L_{\mathrm{m}}$-Off and $L_{\mathrm{m}}$-On. Figure \fref{fig3}{b}
displays the magnitudes of extreme coupling coefficients with $L_{\mathrm{m}}$-Off
and $L_{\mathrm{m}}$-On. As the design goal at the operating frequency
of \SI{6}{\GHz}, the two extreme coupling coefficients have the same
magnitude, which is just the maximum coupling coefficient $K_{0}$.
This means that the same power-splitting performance can be realized
for $L_{\mathrm{m}}$-Off and $L_{\mathrm{m}}$-On at \SI{6}{\GHz}.
In \figref{fig3}{b}, the two curves separate when the frequency
deviates from \SI{6}{\GHz}, which results from the structural dispersions
of different $L_{\mathrm{m}}$. It is clear that the $L_{\mathrm{m}}$-Off
coupling coefficient experiences a stronger dispersion, which will
bring a more severe restriction on the operating bandwidth.

\section{Modulated directional coupler\label{secIV}}

\begin{figure}[ht]
\centering \includegraphics{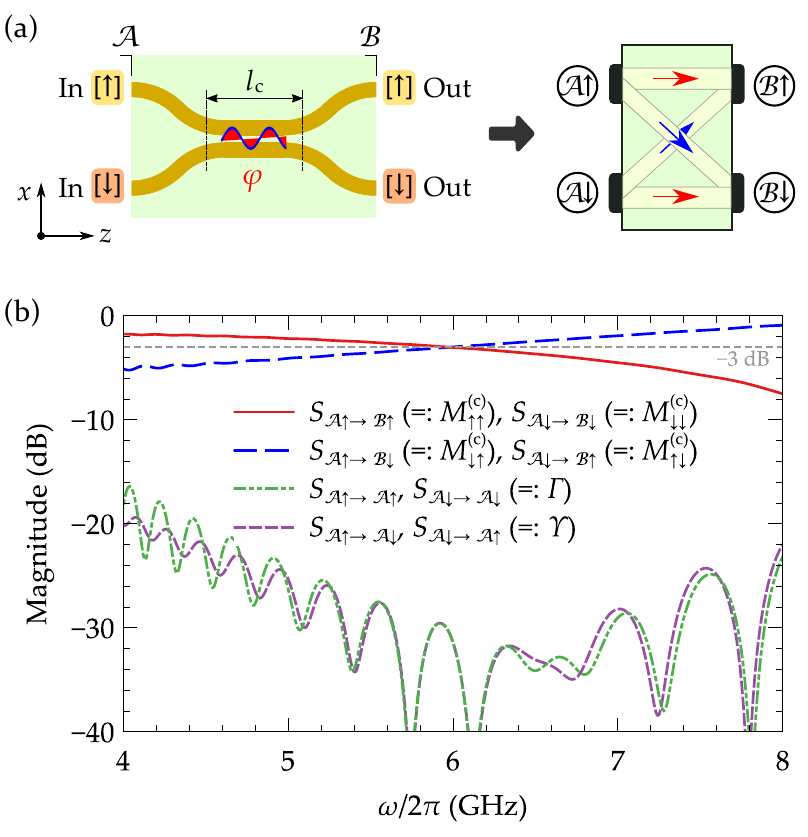} \caption{The 3-dB forward-wave directional coupler. (a) Sketch of a directional
coupler and the reduced four-port network. (b) Scattering parameters
of 3-dB coupler with $l_{\mathrm{c}}=37p$ for operating frequency
of \SI{6}{\GHz}. In the calculations, we adopt the same values of
$\{L_{\mathrm{R}},C_{\mathrm{R}}\}$ and $\{L_{\mathrm{L}},C_{\mathrm{L}}\}$
as these used in \figref{fig3}{a} associated with $C_{\mathrm{m}}=\SI{20}{\fF}$
for the odd-mode and $L_{\mathrm{m}}=\SI{0.5}{\pico\henry}$ (i.e.,
$L_{\mathrm{m}}$-Off) for the even-mode. \label{fig4}}
\end{figure}

\subsection{The 3-dB forward-wave coupler}

To obtain the response of coupled lines, a general approach with the
$ABCD$-matrix for TLs is adopted \citep{Caloz2004ITMTT,Mmongia2007Book}.
First, the $ABCD$-matrix of a unit cell (in a symmetric form) for
an even-line is computed as

\begin{equation}
\left[\begin{array}{cc}
A_{\mathrm{e}} & B_{\mathrm{e}}\\
C_{\mathrm{e}} & D_{\mathrm{e}}
\end{array}\right]{\displaystyle =\left[\begin{array}{cc}
1+Z_{\mathrm{even}}Y_{\mathrm{even}}/2 & Z_{\mathrm{even}}+Z_{\mathrm{even}}^{2}Y_{\mathrm{even}}/4\\
Y_{\mathrm{even}} & 1+Z_{\mathrm{even}}Y_{\mathrm{even}}/2
\end{array}\right]}.
\end{equation}
Then, the $ABCD$-matrix of an $N$-cell even-line can be deduced

\begin{equation}
\left[\begin{array}{cc}
A_{\mathrm{e}N} & B_{\mathrm{e}N}\\
C_{\mathrm{e}N} & D_{\mathrm{e}N}
\end{array}\right]=\left[\begin{array}{cc}
A_{\mathrm{e}} & B_{\mathrm{e}}\\
C_{\mathrm{e}} & D_{\mathrm{e}}
\end{array}\right]^{N}.
\end{equation}
Once the $ABCD$-matrix for an $N$-cell TL has been established,
the corresponding scattering matrix $\mathbf{S}^{[\mathrm{e}]}$of
the $N$-cell even-line for terminations of impedance $Z\mathrm{_{c}}$
can be computed as

\begin{align}
S_{\mathcal{AA}}^{[\mathrm{e}]} & =\left(A_{\mathrm{e}N}+B_{\mathrm{e}N}/Z\mathrm{_{c}}-C_{\mathrm{e}N}Z\mathrm{_{c}}-D_{\mathrm{e}N}\right)/E,\\
S_{\mathcal{AB}}^{[\mathrm{e}]} & =2\left(A_{\mathrm{e}N}D_{\mathrm{e}N}-B_{\mathrm{e}N}C_{\mathrm{e}N}\right)/E,\\
S_{\mathcal{BA}}^{[\mathrm{e}]} & =2/E,\\
S_{\mathcal{BB}}^{[\mathrm{e}]} & =\left(-A_{\mathrm{e}N}+B_{\mathrm{e}N}/Z\mathrm{_{c}}-C_{\mathrm{e}N}Z\mathrm{_{c}}+D_{\mathrm{e}N}\right)/E,
\end{align}
with 

\[
E=A_{\mathrm{e}N}+B_{\mathrm{e}N}/Z\mathrm{_{c}}+C_{\mathrm{e}N}Z\mathrm{_{c}}+D_{\mathrm{e}N}
\]
where the subscripts $\mathcal{A}$ and $\mathcal{B}$ refer to the
two port-related end planes of the calculated $N$-cell segment of
a TL. In the same manner, the scattering matrix $\mathbf{S}^{[\mathrm{o}]}$
of the $N$-cell odd-line can be computed. 

We next consider the response of an $N$-cell CRLH coupler, which
is schematically illustrated in \figref{fig4}{a}. According to
the coupled-mode analyses \citep{Caloz2004ITMTT,Mmongia2007Book},
the $2\times2$ scattering matrix $\mathbf{M}^{(\mathrm{c})}\equiv[M_{\uparrow\uparrow}^{(\mathrm{c})},M_{\uparrow\downarrow}^{(\mathrm{c})};M_{\downarrow\uparrow}^{(\mathrm{c})},M_{\downarrow\downarrow}^{(\mathrm{c})}]$
of a directional coupler can be deduced from the scattering matrices
of the even/odd lines using the following formulas:

\begin{align}
M_{\uparrow\uparrow}^{(\mathrm{c})} & \equiv S_{\mathcal{A}\uparrow\rightarrow\mathcal{B}\uparrow}={\displaystyle (S_{\mathcal{BA}}^{[\mathrm{e}]}+S_{\mathcal{BA}}^{[\mathrm{o}]})/2},\label{eqMc11.static}\\
M_{\downarrow\uparrow}^{(\mathrm{c})} & \equiv S_{\mathcal{A}\uparrow\rightarrow\mathcal{B}\downarrow}={\displaystyle (S_{\mathcal{BA}}^{[\mathrm{e}]}-S_{\mathcal{BA}}^{[\mathrm{o}]})/2},\label{eqMc21.static}\\
M_{\uparrow\downarrow}^{(\mathrm{c})} & \equiv S_{\mathcal{A}\downarrow\rightarrow\mathcal{B}\uparrow}={\displaystyle M_{\downarrow\uparrow}^{(\mathrm{c})}},\label{eqMc12.static}\\
M_{\downarrow\downarrow}^{(\mathrm{c})} & \equiv S_{\mathcal{A}\downarrow\rightarrow\mathcal{B}\downarrow}={\displaystyle M_{\uparrow\uparrow}^{(\mathrm{c})}}.\label{eqMc22.static}
\end{align}
On the other hand, the self- and cross-reflections can also be obtained
using the following expressions:

\begin{align}
\varGamma & \equiv S_{\mathcal{A}\uparrow\rightarrow\mathcal{A}\uparrow}=S_{\mathcal{A}\downarrow\rightarrow\mathcal{A}\downarrow}={\displaystyle (S_{\mathcal{AA}}^{[\mathrm{e}]}+S_{\mathcal{AA}}^{[\mathrm{o}]})/2,}\label{eqGamma}\\
\varUpsilon & \equiv S_{\mathcal{A}\uparrow\rightarrow\mathcal{A}\downarrow}=S_{\mathcal{A}\downarrow\rightarrow\mathcal{A}\uparrow}={\displaystyle (S_{\mathcal{AA}}^{[\mathrm{e}]}-S_{\mathcal{AA}}^{[\mathrm{o}]})/2.}\label{eqUpsilon}
\end{align}

In our design shown in \figref{fig1}{a}, two 3-dB forward-wave
couplers are required to realize the nonreciprocal transmission. To
achieve the 3-dB coupling, $|K_{0}l_{\mathrm{c}}|$ needs to equal
$\pi/4$. Using the magnitude of coupling coefficient at \SI{6}{\GHz}
shown in \figref{fig3}{b}, we obtain a coupling length $l_{\mathrm{c}}$
of $37p$. Since $L_{\mathrm{m}}$-Off induces a stronger dispersion
and causes a more severe restriction on the operating bandwidth, in
what follows, we will focus on behaviors of the 3-dB coupler with
$L_{\mathrm{m}}$-Off. Using equations (\ref{eqMc11.static}--\ref{eqUpsilon}),
we calculate the scattering parameters of the coupler at $L_{\mathrm{m}}$-Off
with terminals of the corresponding uncoupled CRLH TL. The results
are shown in \figref{fig4}{b}, which indicate that the 3-dB power
splittings of the forward beams are achieved at \SI{6}{\GHz}. Meanwhile,
the magnitudes of self- and cross-reflections are very low at around
\SI{6}{\GHz}. For comparison, the responses of $L_{\mathrm{m}}$-On
are shown in Appendix C, which denote that the same 3-dB power splittings
are realized at \SI{6}{\GHz} but with a weaker dispersion.

\subsection{Dynamic modulation of coupling coefficient}

\begin{figure}[ht]
\centering \includegraphics{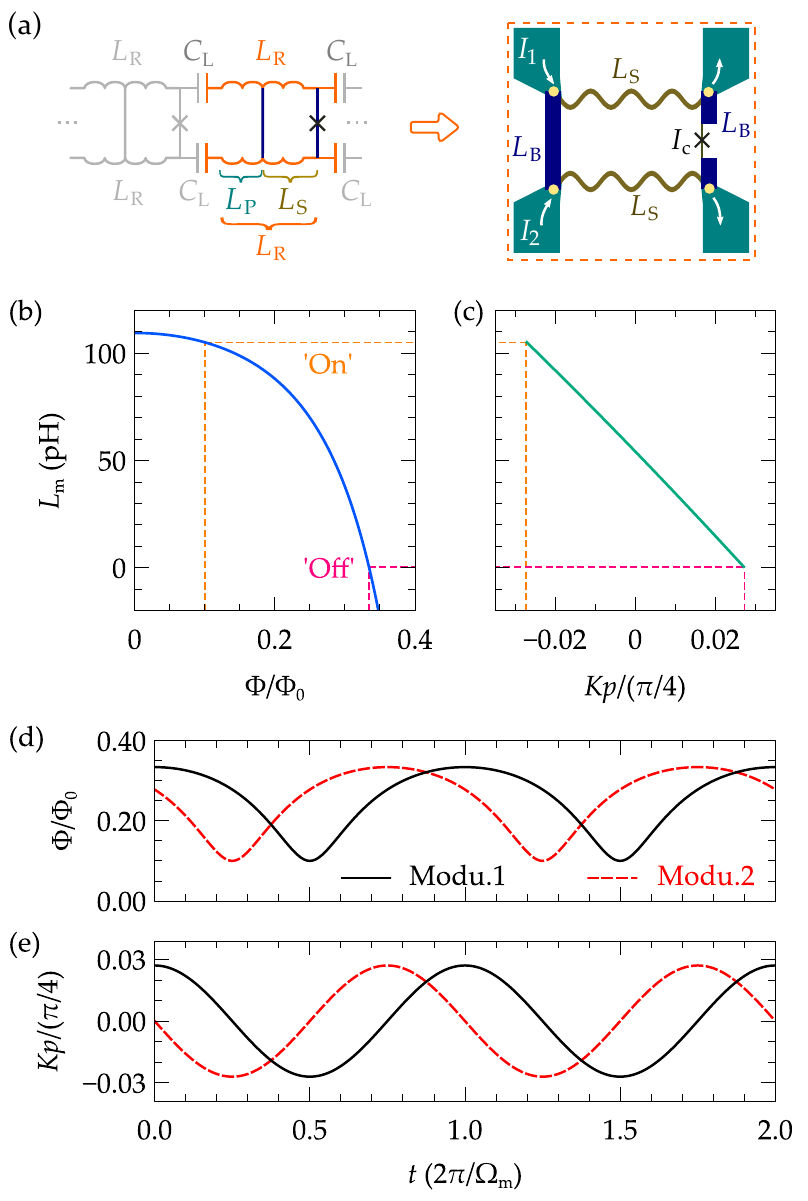} \caption{Coupling modulation via an rf-SQUID. (a) Circuit model for two coupled
lines using an rf-SQUID. Only $L_{\mathrm{R}}$ and $C_{\mathrm{L}}$
of the coupled-line shown in \figref{fig2}{b} are displayed here
for simplicity. The colored drawings illustrate a unit cell of the
coupled-line. The right inset shows a possible design sketch. (b)
The $\Phi$ dependence of $L_{\mathrm{m}}$. (c) Relationship of $K$
with $L_{\mathrm{m}}$. (d) Required time-varying $\Phi$ and (e)
the corresponding modulated $K$ with two different initial phases
of $0$ and $\pi/2$. \label{fig5}}
\end{figure}

To dynamically modulate the coupling coefficient, a radio-frequency
(rf) superconducting quantum interference device (SQUID) is introduced
between two coupled lines. As shown in \figref{fig5}{a}, an rf-SQUID
is coupled galvanically to right-handed segments of two CRLH TLs in
each unit cell. We assume that in a unit cell, the contribution to
$L_{\mathrm{R}}$ can be split into two parts: one is shared with
the rf-SQUID loop and denoted $L_{\mathrm{S}}$; the left, including
parasitic inductance, is depicted by $L_{\mathrm{P}}$, which means
that $L_{\mathrm{R}}=L_{\mathrm{S}}+L_{\mathrm{P}}$. In our calculations,
with $L_{\mathrm{R}}$ equal to \SI{300}{\pico\henry}, a reasonable
and achievable splitting is $\{L_{\mathrm{S}},L_{\mathrm{P}}\}=\{\SI{200}{\pico\henry},\SI{100}{\pico\henry}\}$.
Then, from the perspective of the rf-SQUID, the related superconducting
loop has a geometric inductance $L_{\mathrm{G}}=2L_{\mathrm{S}}+2L_{\mathrm{B}}$,
which is interrupted by a Josephson junction with critical current
$I_{\mathrm{c}}$, as shown in the inset of \figref{fig5}{a}. With
these settings and definitions, we can express the effective SQUID
inductance as

\begin{equation}
L_{\mathrm{rf}}(\Phi)=L_{\mathrm{G}}\frac{1+\beta_{\mathrm{L}}\cos(2\pi\Phi/\Phi_{0})}{\beta_{\mathrm{L}}\cos(2\pi\Phi/\Phi_{0})},\label{eqSQUID.induct}
\end{equation}
where $\Phi_{0}$ is the flux quantum, $\beta_{\mathrm{L}}=2\pi I_{\mathrm{c}}L_{\mathrm{G}}/\Phi_{0}$
is the screening parameter, and $\Phi$ is the flux threading the
SQUID loop. With these expressions and an assumption of $\beta_{\mathrm{L}}<1$,
the effective mutual inductance can be estimated as

\begin{equation}
L_{\mathrm{m}}(\Phi)\simeq L_{\mathrm{dir}}+L_{\mathrm{S}}^{2}/L_{\mathrm{rf}}(\Phi),\label{eqMutual.induct}
\end{equation}
where $L_{\mathrm{dir}}$ expresses a flux-independent mutual inductance
for direct inductive coupling between two lines, and the second term
describes the effective second-order mutual inductance mediated by
the SQUID \citep{Peropadre2013PRB,Wulschner2016EQT}. For the setting
of $\{L_{\mathrm{S}},L_{\mathrm{B}}\}=\{\SI{200}{\pico\henry},\SI{40}{\pico\henry}\}$,
a reasonable value of \SI{70}{\pico\henry} for $L_{\mathrm{dir}}$
can be achieved with a careful design. Then, by using Eqs.\,(\ref{eqSQUID.induct})
and (\ref{eqMutual.induct}), $L_{\mathrm{m}}$ as a function of $\Phi$
is calculated with $\beta_{\mathrm{L}}=0.9$ (i.e., $I_{\mathrm{c}}\simeq\SI{617}{\nA}$)
and plotted in \figref{fig5}{b}. Meanwhile, the $L_{\mathrm{m}}$
dependence of $K$ can be deduced from the results of the analyses
described in Section \ref{secIII}-B and is shown in \figref{fig5}{c}.
The connection between $K$ and $\Phi$ is made on the basis of \figref{fig5}{c}
and \figref{fig5}{b}, which implies that $K$ as a function of
$\Phi$ is obtained. On the basis of such a connection, a temporal
modulation of $K$ can be realized by applying time-varying $\Phi$,
which is easy to implement with a local bias. Here, we only consider
the modulation of $L_{\mathrm{m}}$ ranging from $L_{\mathrm{m}}$-Off
to $L_{\mathrm{m}}$-On as marked in \figref{fig5}{b}. Figures
\fref{fig5}{d} and \fref{fig5}{e} show the required time-varying
$\Phi$ and the corresponding $K$. In \figref{fig5}{e}, the two
modulation curves of $K$ versus time are just the required form in
Eq.\,(\ref{eqModu.coupling}) with different initial phases of $0$
and $\pi/2$. As the explanation given in Appendix A, the required
modulation frequency $\Omega_{\mathrm{m}}$ is about tens of megahertz,
which is not a problem in the presence of local bias.

\section{Nonreciprocal responses\label{secV}}

\begin{figure}[ht]
\centering \includegraphics[width=8.2cm]{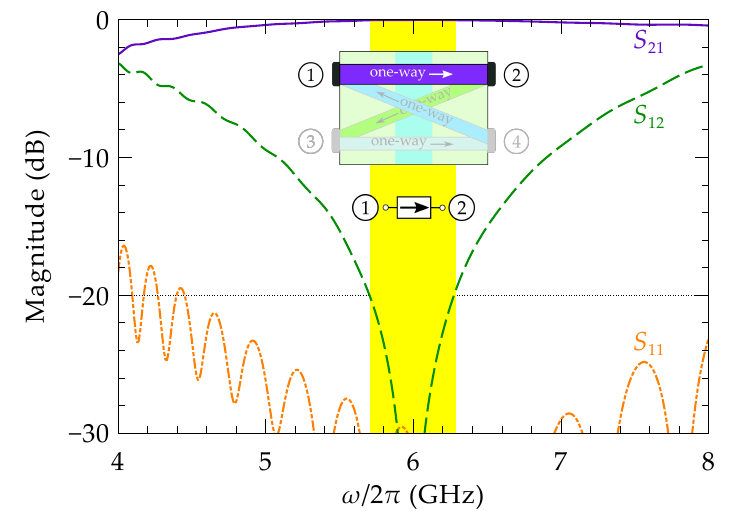} \caption{Nonreciprocal isolation response between port-1 and port-2 of the
structure. A bandwidth of \SI{580}{\MHz} can be achieved with the
isolation over \SI{20}{\dB} at around \SI{6}{\GHz}, where the reflection
is lower than \SI{-29}{\dB}.\label{fig6}}
\end{figure}

After we obtain the implementation of the modulated 3-dB coupler,
the microwave responses of the entire structure with CRLH TLs can
be deduced. Note that the scattering parameters for the coupler in
Eqs.\,(\ref{eqMc11.static}--\ref{eqMc22.static}) are given under
the condition without the modulation of the coupling coefficient.
As described in Section \ref{secII}, when the coupling modulation
expressed in Eq.\,(\ref{eqModu.coupling}) is applied, the scattering
matrix $\mathbf{M}^{(\mathrm{c})}$ of the coupled-line is revised
by the modulation initial phase $\phi$ as

\begin{equation}
\mathbf{M}^{(\mathrm{c})}(\phi)=\left[\begin{array}{cc}
M_{\uparrow\uparrow}^{(\mathrm{c})} & \mathrm{e}^{\mathrm{-j}\phi}M_{\uparrow\downarrow}^{(\mathrm{c})}\\
\mathrm{e}^{\mathrm{j}\phi}M_{\downarrow\uparrow}^{(\mathrm{c})} & M_{\downarrow\downarrow}^{(\mathrm{c})}
\end{array}\right].\label{eqScatM.coupler.CRLH}
\end{equation}
For stage-II, the two CRLH TLs are uncoupled but with different propagation
lengths. The corresponding scattering matrix can be written as

\begin{equation}
\mathbf{M}^{(\mathrm{b})}(\Delta\theta)=\left[\begin{array}{cc}
S_{\mathcal{CB}}^{[\mathrm{\uparrow}]}(\theta_{0}+\Delta\theta) & 0\\
0 & S_{\mathcal{CB}}^{[\mathrm{\downarrow}]}(\theta_{0})
\end{array}\right],\label{eqScatM.phaseshift.CRLH}
\end{equation}
where $\theta_{0}$ is trivial and denotes the common phase for the
two CRLH TLs, which also refers to the operating frequency of \SI{6}{\GHz}.
The scattering parameters $S_{\mathcal{CB}}^{[\mathrm{\uparrow}]}$
and $S_{\mathcal{CB}}^{[\mathrm{\downarrow}]}$ are calculated by
the approach described in Section \ref{secIV}-A with both $L_{\mathrm{m}}$
and $C_{\mathrm{m}}$ equaling zero. By substituting expressions (\ref{eqScatM.coupler.CRLH})
and (\ref{eqScatM.phaseshift.CRLH}) into Eqs.\,(\ref{eqScatM.Forward})
and (\ref{eqScatM.Backward}), we can obtain the scattering matrices
$\mathbf{M}^{(\Rightarrow)}$ and $\mathbf{M}^{(\Leftarrow)}$ for
the entire structure based on CRLH TLs. Here, we redefine each element
as 

\begin{align}
\mathbf{M}^{(\Rightarrow)} & \equiv\left[\begin{array}{cc}
M_{\uparrow\uparrow}^{(\mathrm{\Rightarrow})} & M_{\uparrow\downarrow}^{(\mathrm{\Rightarrow})}\\
M_{\downarrow\uparrow}^{(\mathrm{\Rightarrow})} & M_{\downarrow\downarrow}^{(\mathrm{\Rightarrow})}
\end{array}\right]=\left[\begin{array}{cc}
M_{\mathcal{A}\uparrow\rightarrow\mathcal{D}\uparrow} & M_{\mathcal{A}\downarrow\rightarrow\mathcal{D}\uparrow}\\
M_{\mathcal{A}\uparrow\rightarrow\mathcal{D}\downarrow} & M_{\mathcal{A}\downarrow\rightarrow\mathcal{D}\downarrow}
\end{array}\right],\label{eqScatM.Forward.CRLH}
\end{align}

\begin{align}
\mathbf{M}^{(\mathrm{\Leftarrow})} & \equiv\left[\begin{array}{cc}
M_{\uparrow\uparrow}^{(\Leftarrow)} & M_{\uparrow\downarrow}^{(\Leftarrow)}\\
M_{\downarrow\uparrow}^{(\mathrm{\Leftarrow})} & M_{\downarrow\downarrow}^{(\Leftarrow)}
\end{array}\right]=\left[\begin{array}{cc}
M_{\mathcal{D}\uparrow\rightarrow\mathcal{A}\uparrow} & M_{\mathcal{D}\downarrow\rightarrow\mathcal{A}\uparrow}\\
M_{\mathcal{D}\uparrow\rightarrow\mathcal{A}\downarrow} & M_{\mathcal{D}\downarrow\rightarrow\mathcal{A}\downarrow}
\end{array}\right].\label{eqScatM.Backward.CRLH}
\end{align}
By using these redefined scattering parameters and labeling the four
ports of the entire structure as port-1 to port-4 {[}see \figref{fig1}{a}{]},
we can obtain the total scattering matrix as

\begin{equation}
\mathbf{S}^{(\mathrm{tot)}}=\left[\begin{array}{cccc}
\varGamma & M_{\uparrow\uparrow}^{(\Leftarrow)} & \varUpsilon & M_{\uparrow\downarrow}^{(\Leftarrow)}\\
M_{\uparrow\uparrow}^{(\Rightarrow)} & \varGamma & M_{\uparrow\downarrow}^{(\Rightarrow)} & \varUpsilon\\
\varUpsilon & M_{\downarrow\uparrow}^{(\Leftarrow)} & \varGamma & M_{\downarrow\downarrow}^{(\Leftarrow)}\\
M_{\downarrow\uparrow}^{(\Rightarrow)} & \varUpsilon & M_{\downarrow\downarrow}^{(\Rightarrow)} & \varGamma
\end{array}\right],\label{eqSactS.tot.CRLH}
\end{equation}
where the self-reflection for each port and the cross-reflection between
ports located on the same side appear in the form of Eqs.\,(\ref{eqGamma})
and (\ref{eqUpsilon}), respectively. The reason for using $\varGamma$
and $\varUpsilon$ as estimations for the reflections is that the
cascade-induced inter-reflections are very small for waveguide-based
structures. As a comparison, the exact total scattering matrix is
deduced by a full-network analysis and shown in Appendix D. 

\begin{figure}
\centering \includegraphics[width=8.2cm]{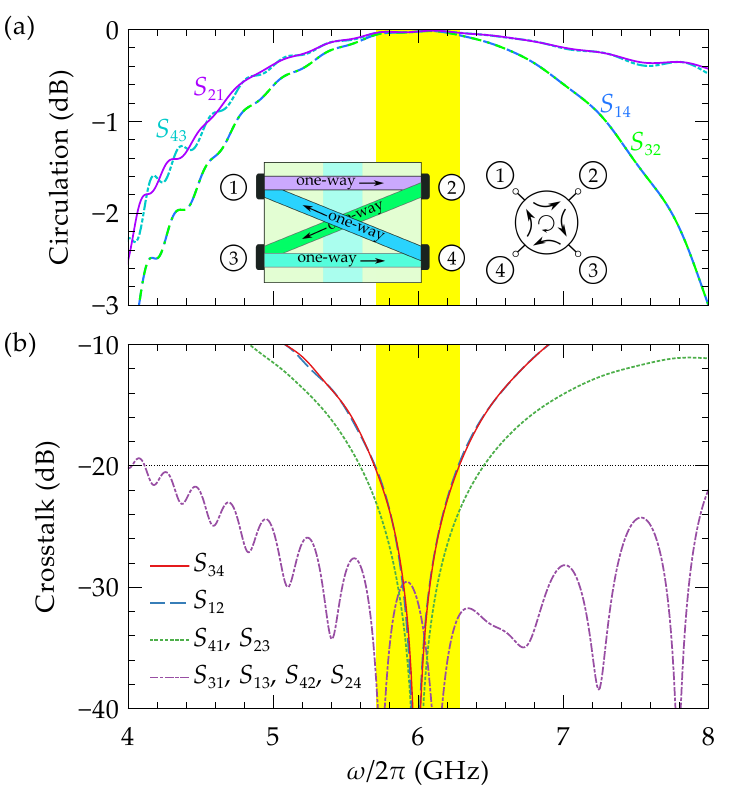} \caption{Nonreciprocal responses of (a) circulation and (b) crosstalk of the
entire four-port structure. A bandwidth of \SI{580}{\MHz} can be
achieved with the isolation over \SI{20}{\dB} at around \SI{6}{\GHz},
where the crosstalks are lower than \SI{-29}{\dB}.\label{fig7}}
\end{figure}

From the results of the analyses in Section \ref{secII}, realizing
nonreciprocity requires a specific setting on modulation phases in
stage-I(-III) and phase shifts in stage-II. Here, $\{\phi_{1},\Delta\theta,\phi_{2}\}=\{0,\pi/2,\pi/2\}$
is adopted again and substituted into equations (\ref{eqScatM.coupler.CRLH})
and (\ref{eqScatM.phaseshift.CRLH}). Note that, because there is
dispersion of the whole system, all the settings of frequency-dependent
parameters refer to our designed operation frequency of \SI{6}{\GHz}.
Note also that $\Delta\theta$ equaling $\pi/2$ requires 2.2 cells
with current design parameters for the uncoupled CRLH TLs, but an
integer multiple of cells can be achieved by slightly adjusting design
parameters in stage-II. After some calculations, the total scattering
parameters can be extracted from Eq.\,(\ref{eqSactS.tot.CRLH}).
Here, we demonstrate its nonreciprocal isolation between two ports
and nonreciprocal circulations of four ports. To give a clear demonstration,
a $2\times2$ scattering matrix related to port-1 and port-2 is extracted
and defined as
\begin{equation}
\mathbf{S}^{(\mathrm{1\Leftrightarrow2)}}\equiv\left[\begin{array}{cc}
S_{11} & S_{12}\\
S_{21} & S_{22}
\end{array}\right]=\left[\begin{array}{cc}
\varGamma & M_{\uparrow\uparrow}^{(\Leftarrow)}\\
M_{\uparrow\uparrow}^{(\Rightarrow)} & \varGamma
\end{array}\right].\label{eqScatS.Iso.CRLH}
\end{equation}

Figure \ref{fig6} shows the scattering parameters between port-1
and port-2. The results indicate that isolation over \SI{20}{\dB}
for transmissions between port-1 and port-2 can be obtained with a
bandwidth of \SI{580}{\MHz} at around \SI{6}{\GHz}. Meanwhile, the
absolute magnitudes of $S_{21}$ and $S_{11}$ show that the insertion
loss and reflection are very small. These merits indicate that a well-performing
isolator is realized by the designed structure with the CRLH TLs.
Furthermore, the full scattering parameters for all four ports are
displayed in Fig.\,\ref{fig7}. A clear nonreciprocal circulation
from port-1 to port-4 is shown in \figref{fig7}{a} and all other
crosstalk channels remain low-magnitude transmissions, as illustrated
in \figref{fig7}{b}. The results of Fig.\,\ref{fig7} represent
that a circulator can also be obtained with the same structure. It
should be noted that, strictly speaking, this kind of circulator is
not a common circulator, because a frequency shift of $\Omega_{\mathrm{m}}$
is introduced when waves are transmitted between ports attached to
different TLs. Specifically, the frequency shift occurs in the transmission
of $S_{32}$ and $S_{14}$, resulting from\textcolor{red}{{} }the interband
transition between mode-{[}\textuparrow {]} and mode-{[}\textdownarrow {]}
induced by the dynamic modulation of the coupling coefficient (see
Appendix A). Such a nonreciprocal circulation can be classified as
the nonreciprocity created in the frequency space \citep{Lecocq2017PRA,Sliwa2015PRX,Abdelsalam2020O}.
A detailed interpretation is shown in Appendix A. However, such a
nonreciprocal circulation can be used for applications concerned only
with transfers of power. 

\begin{figure*}
\centering \includegraphics{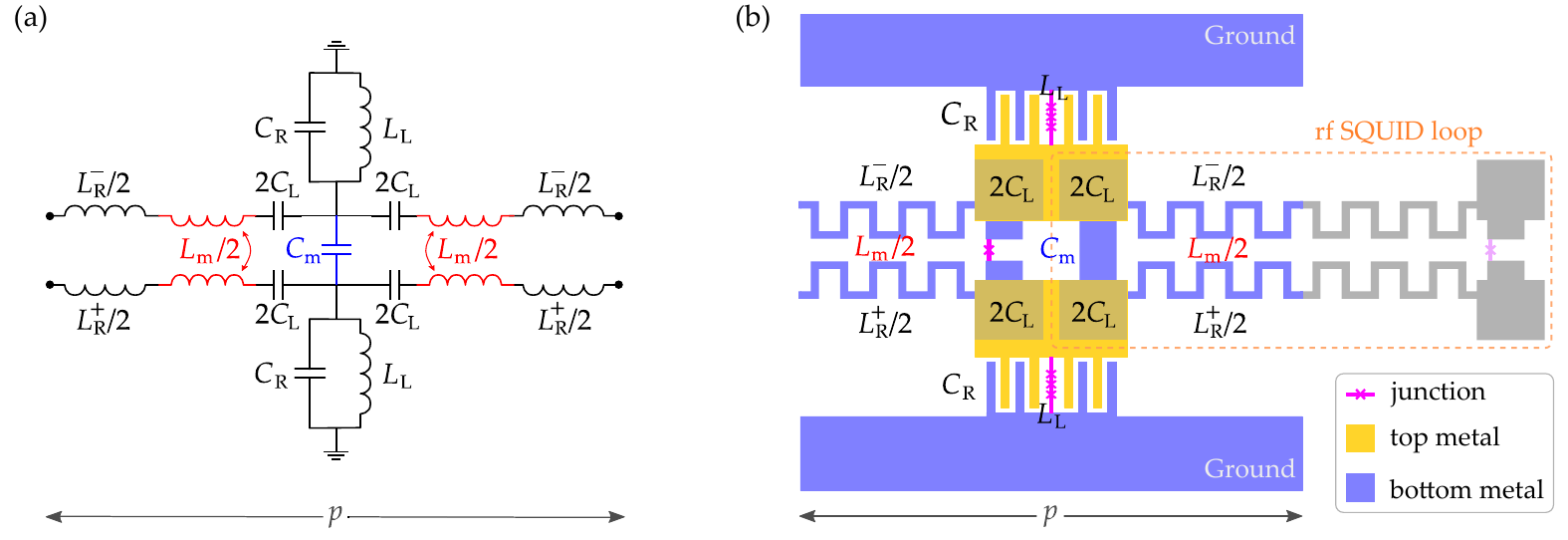} \caption{Physical implementation for the coupler design. (a) Circuit model
for a symmetric unit cell of CRLH coupled-line. To achieve a slight
asymmetry between two lines, small negative and positive $L_{\mathrm{R}}$-shifts
are introduced in the upper and lower lines, respectively. (b) Design
layout for a symmetric unit cell of CRLH coupled-line and the rf-SQUID
loop for modulation of coupling coefficient. \label{fig8}}
\end{figure*}

\section{discussion\label{secVI}}

To make the design into a reality from the circuit models, here, we
propose to fabricate the CRLH TLs by a standard fabrication process
for superconducting circuits. In general, to make the forward and
backward waves see identical input impedance, a symmetric design of
the unit cell of the coupled CRLH TLs is adopted and shown in \figref{fig8}{a}
{[}compare with \figref{fig2}{b}{]}. By recalling the design parameters
$\{L_{\mathrm{R}},C_{\mathrm{R}}\}=\{\SI{300}{\pico\henry},\SI{150}{\fF}\}$,
these two contributions are realized with the common meander line
and interdigital capacitor. To achieve an efficient transition between
mode-$[\uparrow]$ and mode-$[\downarrow]$ under dynamic modulation,
small $L_{\mathrm{R}}$-shifts can be implemented by slightly adjusting
the width of meander lines (see Appendix A). The left-handed elements
of $\{L_{\mathrm{L}},C_{\mathrm{L}}\}=\{\SI{1400}{\pico\henry},\SI{560}{\fF}\}$
can be fabricated using a Josephson junction array and a parallel
capacitor, in order to pursue a compact size and reduce the parasitic
inductance and capacitance. Note that the parasitic inductance and
capacitance should be considered and contributed to $\{L_{\mathrm{R}},C_{\mathrm{R}}\}$.
A careful design layout is needed to meet the required $C_{\mathrm{m}}=\SI{20}{\fF}$,
and $L_{\mathrm{m}}=\SI{105}{\pico\henry}$ for the $L_{\mathrm{m}}$-On.
The $L_{\mathrm{m}}$-Off and modulation of $L_{\mathrm{m}}$ are
realized by constructing an rf-SQUID loop. Figure \fref{fig8}{b}
displays a possible design layout for a unit cell, where the rf-SQUID
loop is also outlined. A single local-bias line for all cells used
for modulating the rf-SQUID loops can be designed and fabricated using
airbridge technology \citep{Chen2014APL,Mukai2020NJP}, which is not
illustrated in \figref{fig8}{b}. The uncoupled CRLH TLs are easy
to achieve by spatially separating the two TLs to cease their coupling.
Importantly, the length of one unit cell less than \SI{300}{\um}
is possible for these design parameters. Considering the cell numbers
of 37 for each coupler in stage-I (stage-III) and (at least) 2 for
the phase shift in stage-II, the estimated length for the entire structure
is about \SI{23}{\mm}, which is an acceptable size for a superconducting
circuit chip. This means that a compact nonreciprocal isolator or
circulator can be realized with our design on a microwave circuit
chip.

In this work, we designed and demonstrated an isolator (circulator)
with a bandwidth of \SI{580}{\MHz} at around \SI{6}{\GHz}. However,
these two metrics can be improved further by adjusting the design
parameters. The operating frequency is readily changed by forming
correspondingly responded couplers in stage-I (stage-III) and phase
shifts in stage-II. A wider bandwidth is also possible; one solution
is using a weaker dispersed coupled-line, but at a cost of the footprint
of the entire structure. Another advanced approach is designing a
super unit cell with a more linear dispersion curve (i.e., a smaller
group-velocity dispersion), where a delicate design is required \citep{OBrien2014PRL,Planat2020PRX}.
Moreover, the reflection (i.e., return loss) can be further reduced
by adopting a tapered structure to more closely match the \SI{50}{\ohm}
line. Here, we demonstrated the design by the modulations of inductance
or permeability in the couplers, but a similar functionality can also
be implemented by modulating capacitance or permittivity in the other
types of circuit \citep{Nagulu2020NE,Dinc2017NC}. As a spatiotemporal
modulation-based nonreciprocal design, compared with the operating
frequency, the modulation frequency is usually very low. Thus, the
modulation wave can be simply filtered by spectral and spatial designs.
Thus, in our scheme, in comparison with the general nonlinearity-based
nonreciprocity, there is no strong microwave pumping, whose frequency
is comparable to the operating frequency. This feature is also of
benefit to its applications in quantum measurements.

\section{Conclusion\label{secVII}}

In summary, we proposed a design of a magnetic-free broadband isolator
or circulator with standard superconducting circuits. The designed
structure consists of three stages, where two directional couplers
are connected by inserted two uncoupled lines with different phase
shifts. The two couplers require temporal modulations of the coupling
coefficients with different initial phases. The entire structure can
be fabricated using superconducting lines and Josephson junctions
with a total line length of \SI{23}{\mm}. In this work, the calculated
results show that the nonreciprocal isolation over \SI{20}{\dB} with
the bandwidth of \SI{580}{\MHz} is realized at around the frequency
of \SI{6}{\GHz}. Nonreciprocal circulations for microwave powers
can also be realized with this structure. Moreover, these figures
of merit can be further enhanced through the structural optimization
design. Such an isolator or circulator will be very useful for cryogenic
microwave connections or measurements in quantum integrated circuits.
\begin{acknowledgments}
The authors acknowledge useful discussions with Rui Wang and Iulia
Zotova. This work was supported by JST {[}Moonshot R\&D{]}{[}Grant
Number JPMJMS2067{]} and CREST, JST (Grant No. JPMJCR1676). This article
was partly based on results obtained from a project, JPNP16007, commissioned
by the New Energy and Industrial Technology Development Organization
(NEDO). 

\vspace{0.2cm}
\emph{Note added.}---Recently, we became aware of another work using
a similar method to demonstrate traveling-wave nonreciprocity\textcolor{red}{{}
}\citep{Naghiloo2021aX2103.07793}.
\end{acknowledgments}

%\widetext 
%\clearpage

%%%%%%%%%%%%%%%%%%%%%%%%%%%%%%%%%%%%%%%%%%%%%%%%%%%%%%%%%%%%%%%%%%%%%%%
%%%%%%%%%%%%%%%%%%%%%%%%%%%%%%%%%%%%%%%%%%%%%%%%%%%%%%%%%%%%%%%%%%%%%%%
\renewcommand{\theequation}{A\arabic{equation}} 
\setcounter{equation}{0} 

\section*{Appendix A: Coupled-mode theory}

\begin{figure}
\centering 

\includegraphics[width=8.5cm]{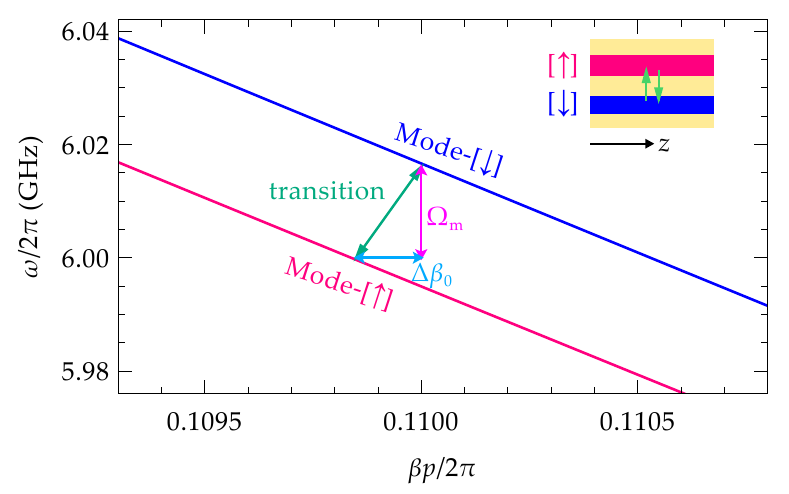} \caption{Interband transition or dynamical coupling between mode-$[\uparrow]$
and mode-$[\downarrow]$ of a dual-waveguide structure. \label{fig9}}
\end{figure}

We start by considering the photon transition process in a dual-waveguide
system. The uncoupled dual-waveguide structure is represented by a
permeability distribution $\text{\ensuremath{\mu}}_{\mathrm{s}}(\mathbf{r}_{\perp})$
{[}$\mathbf{r}_{\perp}=(x,y)${]}, which is time-independent (i.e.,
static) and uniform along the $z$-direction (see the inset of Fig.\,\ref{fig9}).
Such an uncoupled dual-waveguide structure possesses a band structure,
as shown in Fig.\,\ref{fig9}, with up-located and down-located modes
corresponding to the first and second bands (i.e., mode-$[\uparrow]$
and mode-$[\downarrow]$), respectively. An interband transition between
the two modes with frequencies and wave vectors, $(\omega_{\uparrow},k_{\uparrow})$,
$(\omega_{\downarrow},k_{\downarrow})$, can be induced by modulating
the dual-waveguide structure with an additional time-varying relative
permeability perturbation 

\begin{equation}
\mu(\mathbf{r}_{\perp},t)=\mu_{\mathrm{s}}(\mathbf{r}_{\perp})+\Delta\mu(\mathbf{r}_{\perp},t),\label{eqMuDis}
\end{equation}
with

\begin{equation}
\Delta\mu(\mathbf{r}_{\perp},t)=\Delta\mu_{\mathrm{m}}(\mathbf{r}_{\perp})\cos(\Omega_{\mathrm{m}}t+\phi),\label{eqModuMuDis}
\end{equation}
where $\Delta\mu_{\mathrm{m}}(\mathbf{r}_{\perp})$ is the modulation
amplitude distribution along the cross section, $\Omega_{\mathrm{m}}=\omega_{\downarrow}-\omega_{\uparrow}$
is the modulation frequency, and $\phi$ is the initial phase of the
modulation profile \citep{Yu2009NP,Fang2012PRL,Qin2018PRL}. 

\begin{widetext}

In the modulated dual-waveguide structure, the magnetic field can
be expressed as

\begin{equation}
H(\mathbf{r}_{\perp},z,t)=a_{\uparrow}(z)\psi_{\uparrow}(\mathbf{r}_{\perp})\mathrm{e}^{\mathrm{j}(\omega_{\uparrow}t-\beta_{\uparrow}z)}+a_{\downarrow}(z)\psi_{\downarrow}(\mathbf{r}_{\perp})\mathrm{e}^{\mathrm{j}(\omega_{\downarrow}t-\beta_{\downarrow}z)},\label{eqHfieldInter}
\end{equation}
where $\psi_{\uparrow,\downarrow}(\mathbf{r}_{\perp})$ are the modal
profiles and $a_{\uparrow,\downarrow}(z)$ denote the field amplitudes
of mode-$[\uparrow]$ and mode-$[\downarrow]$. For simplicity, we
have assumed the quasi-TEM modes where the magnetic field has components
majorly along the $\mathbf{r}_{\perp}$-direction. The magnetic field
in Eq.\,(\ref{eqHfieldInter}) satisfies Maxwell's wave equation,

\begin{equation}
{\displaystyle \nabla^{2}H(\mathbf{r}_{\perp},z,t)=\frac{1}{c^{2}}\frac{\partial^{2}}{\partial t^{2}}[\mu(\mathbf{r}_{\perp},t)H(\mathbf{r}_{\perp},z,t)]}.\label{eqHfWaveEqu}
\end{equation}
First, we consider the dual-waveguide structure without modulation
(i.e., static). The magnetic fields satisfy the orthonormalization
condition

\begin{equation}
{\displaystyle \frac{c^{2}\beta_{i}}{2\omega_{i}^{2}}\int_{-\infty}^{\infty}\psi_{i}^{*}(\mathbf{r}_{\perp})\psi_{j}(\mathbf{r}_{\perp})=\delta_{ij}}.\qquad i,j=\uparrow,\downarrow\label{eqEfNormInter}
\end{equation}
With such a normalization, $|a_{i}|^{2}$ denotes the photon number
flux carried by the mode-$[i]$ ($i=\uparrow,\downarrow$). Then,
we consider the modulated dual-waveguide structure. We substitute
Eqs.\,(\ref{eqMuDis}), (\ref{eqModuMuDis}), and (\ref{eqHfieldInter})
into Eq.\,(\ref{eqHfWaveEqu}), 

\begin{equation}
\left(\nabla_{\perp}^{2}+\frac{\partial^{2}}{\partial z^{2}}\right)H(\mathbf{r}_{\perp},z,t)=\frac{\mu_{\mathrm{s}}(\mathbf{r}_{\perp})}{c^{2}}\frac{\partial^{2}}{\partial t^{2}}H(\mathbf{r}_{\perp},z,t)+{\displaystyle \frac{\Delta\mu_{\mathrm{m}}(\mathbf{r}_{\perp})}{c^{2}}\frac{\partial^{2}}{\partial t^{2}}\left[\cos(\Omega t+\phi)H(\mathbf{r}_{\perp},z,t)\right]}.\label{eqEfWaveEqu1.tsInter}
\end{equation}
Then we have

\begin{equation}
\begin{aligned} & \mathrm{e}^{\mathrm{j}(\omega_{\uparrow}t-\beta_{\uparrow}z)}\left(-\mathrm{j}2\beta_{\uparrow}\psi_{\uparrow}\frac{\partial a_{\uparrow}}{\partial z}-a_{\uparrow}\beta_{\uparrow}^{2}\psi_{\uparrow}+a_{\uparrow}\nabla_{\perp}^{2}\psi_{\uparrow}\right)+\mathrm{e}^{\mathrm{j}(\omega_{\downarrow}t-\beta_{\downarrow}z)}\left(-\mathrm{j}2\beta_{\downarrow}\psi_{\downarrow}\frac{\partial a_{\downarrow}}{\partial z}-a_{\downarrow}\beta_{\downarrow}^{2}\psi_{\downarrow}+a_{\downarrow}\nabla_{\perp}^{2}\psi_{\downarrow}\right)\\
 & \qquad=\mathrm{e}^{\mathrm{j}(\omega_{\uparrow}t-\beta_{\uparrow}z)}\left(-\frac{\Delta\mu_{\mathrm{m}}}{2}\left[(\omega_{\uparrow}+\Omega_{\mathrm{m}})^{2}\mathrm{e}^{\mathrm{j}(\Omega_{\mathrm{m}}t+\phi)}+(\omega_{\uparrow}-\Omega_{\mathrm{m}})^{2}\mathrm{e}^{\mathrm{-j}(\Omega_{\mathrm{m}}t+\phi)}\right]-\omega_{\uparrow}^{2}\mu_{\mathrm{s}}\right)\frac{a_{\uparrow}\psi_{\uparrow}}{c^{2}}\\
 & \qquad\quad+\mathrm{e}^{\mathrm{j}(\omega_{\downarrow}t-\beta_{\downarrow}z)}\left(-\frac{\Delta\mu_{\mathrm{m}}}{2}\left[(\omega_{\downarrow}+\Omega_{\mathrm{m}})^{2}\mathrm{e}^{\mathrm{j}(\Omega_{\mathrm{m}}t+\phi)}+(\omega_{\downarrow}-\Omega_{\mathrm{m}})^{2}\mathrm{e}^{\mathrm{-j}(\Omega_{\mathrm{m}}t+\phi)}\right]-\omega_{\downarrow}^{2}\mu_{\mathrm{s}}\right)\frac{a_{\downarrow}\psi_{\downarrow}}{c^{2}},
\end{aligned}
\label{eqEfWaveEqu2.tsInter}
\end{equation}
where we have used the slowly varying envelope approximation and dropped
the terms $\partial^{2}a_{\uparrow,\downarrow}/\partial z^{2}$. Furthermore,
on the basis of the definition of the modes in the unmodulated (i.e.,
static) waveguide and Eq.\,(\ref{eqHfWaveEqu}), we obtain

\begin{equation}
-\beta_{i}^{2}\psi_{i}+\nabla_{\perp}^{2}\psi_{i}=-\omega_{i}^{2}\mu_{\mathrm{s}}\frac{\psi_{i}}{c^{2}},\qquad i=\uparrow,\downarrow
\end{equation}
and by considering rotating wave approximation, we can derive the
coupled-mode equation from Eq.\,(\ref{eqEfWaveEqu2.tsInter}),

\begin{equation}
\begin{aligned} & \mathrm{j}2\beta_{\uparrow}\psi_{\uparrow}\frac{\partial a_{\uparrow}}{\partial z}\mathrm{e}^{\mathrm{j}(\omega_{\uparrow}t-\beta_{\uparrow}z)}+\mathrm{j}2\beta_{\downarrow}\psi_{\downarrow}\frac{\partial a_{\downarrow}}{\partial z}\mathrm{e}^{\mathrm{j}(\omega_{\downarrow}t-\beta_{\downarrow}z)}\\
 & \qquad=\frac{\Delta\mu_{\mathrm{m}}\psi_{\uparrow}}{2c^{2}}(\omega_{\uparrow}+\Omega_{\mathrm{m}})^{2}a_{\uparrow}\mathrm{e}^{\mathrm{j}\left[(\omega_{\uparrow}+\Omega_{\mathrm{m}})t-\beta_{\uparrow}z+\phi\right]}+\frac{\Delta\mu_{\mathrm{m}}\psi_{\downarrow}}{2c^{2}}(\omega_{\downarrow}-\Omega_{\mathrm{m}})^{2}a_{\downarrow}\mathrm{e}^{\mathrm{j}\left[(\omega_{\downarrow}-\Omega_{\mathrm{m}})t-\beta_{\downarrow}z-\phi\right]},
\end{aligned}
\end{equation}
then, we have the equation of motion of the modulated dual-waveguide,

\begin{equation}
{\displaystyle \mathrm{j}\frac{\mathrm{\partial}}{\partial z}\left[\begin{array}{c}
a_{\uparrow}\\
a_{\downarrow}
\end{array}\right]=\left[\begin{array}{cc}
0 & K_{\uparrow\downarrow}\mathrm{e}^{-\mathrm{j}\left(\Delta\beta_{0}z+\phi\right)}\\
K_{\downarrow\uparrow}\mathrm{e}^{\mathrm{j}\left(\Delta\beta_{0}z+\phi\right)} & 0
\end{array}\right]\left[\begin{array}{c}
a_{\uparrow}\\
a_{\downarrow}
\end{array}\right]},\label{eqCME.tsInter}
\end{equation}
or, alternatively, the corresponding Hamiltonian can be given by

\begin{equation}
\mathcal{H}=K_{\uparrow\downarrow}\mathrm{e}^{\mathrm{-j}\left(\Delta\beta_{0}z+\phi\right)}a_{\uparrow}^{\dagger}a_{\downarrow}+K_{\downarrow\uparrow}\mathrm{e}^{\mathrm{j}\left(\Delta\beta_{0}z+\phi\right)}a_{\uparrow}a_{\downarrow}^{\dagger},\label{eqHam.tsInter}
\end{equation}

\noindent where $\Delta\beta_{0}=\beta_{\downarrow}-\beta_{\uparrow}$,
and the coupling coefficient $K_{ij}\:(i\neq j)$ is calculated as

\begin{equation}
K_{ij}=\frac{{\displaystyle \omega_{i}}\iint\mathrm{d}x\mathrm{d}y\Delta\mu_{\mathrm{m}}(\mathbf{r}_{\perp})\psi_{i}^{*}(\mathbf{r}_{\perp})\psi_{j}(\mathbf{r}_{\perp})}{4c^{2}\beta_{i}\iint\mathrm{d}x\mathrm{d}y\psi_{i}^{*}(\mathbf{r}_{\perp})\psi_{i}(\mathbf{r}_{\perp})},
\end{equation}
with the use of the orthonormalization condition of Eq.\,(\ref{eqEfNormInter}),
we obtain

\begin{equation}
K_{\uparrow\downarrow}=K_{\downarrow\uparrow}^{*}={\displaystyle \frac{1}{8}}\iint\mathrm{d}x\mathrm{d}y\Delta\mu_{\mathrm{m}}(\mathbf{r}_{\perp})\psi_{\uparrow}^{*}(\mathbf{r}_{\perp})\psi_{\downarrow}(\mathbf{r}_{\perp}).\label{eqCoupl.rate.expression}
\end{equation}
Generally, a pure real coupling coefficient can be obtained by the
proper selection of initial phases of $\psi_{\uparrow}(\mathbf{r}_{\perp})$
and $\psi_{\downarrow}(\mathbf{r}_{\perp})$ in (\ref{eqCoupl.rate.expression})
and appears in the form 

\begin{equation}
K_{0}\equiv K_{\uparrow\downarrow}=K_{\downarrow\uparrow}.\label{eqRealCoupl.rate}
\end{equation}
The solution to Eq.\,(\ref{eqCME.tsInter}) is

\begin{equation}
\left[\begin{array}{c}
a_{\uparrow}(z)\\
a_{\downarrow}(z)
\end{array}\right]=\left[\begin{array}{cc}
\mathrm{e}^{-\mathrm{j}\frac{\Delta\beta_{\mathrm{0}}z}{2}}{\displaystyle \left\{ \cos\left(\frac{\Delta\beta_{\mathrm{c}}z}{2}\right)+\mathrm{j}\frac{\Delta\beta_{0}}{\Delta\beta_{\mathrm{c}}}\sin\left(\frac{\Delta\beta_{\mathrm{c}}z}{2}\right)\right\} } & \mathrm{-j}\mathrm{e}^{\mathrm{j}\left(\frac{\Delta\beta_{\mathrm{0}}z}{2}+\phi\right)}{\displaystyle \frac{2K_{0}}{\Delta\beta_{\mathrm{c}}}\sin\left(\frac{\Delta\beta_{\mathrm{c}}z}{2}\right)}\\
\mathrm{-j}\mathrm{e}^{\mathrm{-j}\left(\frac{\Delta\beta_{\mathrm{0}}z}{2}+\phi\right)}{\displaystyle \frac{2K_{0}}{\Delta\beta_{\mathrm{c}}}\sin\left(\frac{\Delta\beta_{\mathrm{c}}z}{2}\right)} & \mathrm{e}^{\mathrm{j}\frac{\Delta\beta_{\mathrm{0}}z}{2}}{\displaystyle \left\{ \cos\left(\frac{\Delta\beta_{\mathrm{c}}z}{2}\right)-\mathrm{j}\frac{\Delta\beta_{0}}{\Delta\beta_{\mathrm{c}}}\sin\left(\frac{\Delta\beta_{\mathrm{c}}z}{2}\right)\right\} }
\end{array}\right]\left[\begin{array}{c}
a_{\uparrow}(0)\\
a_{\downarrow}(0)
\end{array}\right],\label{eqSolu_tsInter}
\end{equation}
where

\begin{align}
\Delta\beta_{\mathrm{c}} & =2\sqrt{(\Delta\beta_{0}/2)^{2}+K_{0}^{2}}.\label{eqBetasplitting.coupling}
\end{align}

\end{widetext}

\noindent By combining equations (\ref{eqModuMuDis}), (\ref{eqCoupl.rate.expression}),
and (\ref{eqRealCoupl.rate}), we found that a time-varying coupling
is generated by such a modulation of permeability and expressed as 

\begin{equation}
K(t)=K_{0}\cos(\Omega_{\mathrm{m}}t+\phi).
\end{equation}

\noindent \emph{Phase-matching case.} To achieve an efficient interband
transition, the phase-matching condition should be satisfied. Thus,
we can assume $\Delta\beta_{0}=0$, and Eq.\,(\ref{eqCME.tsInter})
is reduced to

\begin{equation}
{\displaystyle \mathrm{j}\frac{\mathrm{\partial}}{\partial z}\left[\begin{array}{c}
a_{\uparrow}\\
a_{\downarrow}
\end{array}\right]=\left[\begin{array}{cc}
0 & K_{0}\mathrm{e}^{-\mathrm{j}\phi}\\
K_{0}\mathrm{e}^{\mathrm{j}\phi} & 0
\end{array}\right]\left[\begin{array}{c}
a_{\uparrow}\\
a_{\downarrow}
\end{array}\right]},\label{eqCME.tsInter-1}
\end{equation}
and the corresponding Hamiltonian can be expressed as

\begin{equation}
\mathcal{H}=K_{0}\mathrm{e}^{-\mathrm{j}\phi}a_{\uparrow}^{\dagger}a_{\downarrow}+K_{0}\mathrm{e}^{\mathrm{j}\phi}a_{\uparrow}a_{\downarrow}^{\dagger}.\label{eqHam.tsInter-1}
\end{equation}
Correspondingly, the solution to Eq.\,(\ref{eqCME.tsInter-1}) is

\begin{equation}
\left[\begin{array}{c}
a_{\uparrow}(z)\\
a_{\downarrow}(z)
\end{array}\right]=\left[\begin{array}{cc}
{\displaystyle \cos\left(K_{0}z\right)} & -\mathrm{j}\mathrm{e}^{-\mathrm{j}\phi}{\displaystyle \sin\left(K_{0}z\right)}\\
-\mathrm{j}\mathrm{e}^{\mathrm{j}\phi}{\displaystyle \sin\left(K_{0}z\right)} & {\displaystyle \cos\left(K_{0}z\right)}
\end{array}\right]\left[\begin{array}{c}
a_{\uparrow}(0)\\
a_{\downarrow}(0)
\end{array}\right],\label{eqSolu_tsInter-1}
\end{equation}
Importantly, with Eq.\,(\ref{eqBetasplitting.coupling}), there is
$\Delta\beta_{\mathrm{c}}=2K_{0}$, which corresponds to the wavenumber
splitting of two normal modes induced by the coupling. When $\Omega_{\mathrm{m}}=\omega_{\downarrow}-\omega_{\uparrow}$
approaches zero (i.e., $\Omega_{\mathrm{m}}\ll\omega_{\downarrow},\omega_{\uparrow}$),
the two normal modes are known as the even-mode and odd-mode, so there
is the expression

\begin{equation}
K_{0}=\frac{\beta_{\mathrm{even}}-\beta_{\mathrm{odd}}}{2},\label{eqCoupRate.even-odd}
\end{equation}
where $\beta_{\mathrm{even}}$ and $\beta_{\mathrm{odd}}$ are propagation
constants (i.e., wavenumbers) of the even and odd modes, respectively.
For the dual-mode nonreciprocity, a more general description is presented
in Ref.~\citep{Ranzani2014NJP}, where a useful insight is provided
by a geometric picture.

In our design, to simultaneously realize the temporal modulation and
efficient transition, a small frequency difference (i.e., $\Omega_{\mathrm{m}}$)
between mode-$[\uparrow]$ and mode-$[\downarrow]$ is required at
$\Delta\beta_{0}=0$. In our implementation with superconducting circuit
elements, this frequency difference can be readily realized by slightly
tuning the design parameters. For example, at the operation frequency
of \SI{6}{\GHz}, $\Omega_{\mathrm{m}}=\SI{20}{\MHz}$ is achieved
by adding an $L_{\mathrm{R}}$-shift of $-6$ and \SI{+6}{\pico\henry}
for mode-$[\uparrow]$ and mode-{[}$\downarrow${]}, respectively,
as shown in Fig.\,\ref{fig9}. Notice that, when a different $L_{\mathrm{R}}$-shift
is introduced to two waveguides, the coupled modes will not be perfect
even and odd modes, which are called even-like and odd-like modes.
However, these small shifts induce a negligible modification to the
final results, since the magnitude of \SI{+6}{\pico\henry} of the
$L_{\mathrm{R}}$-shift is much smaller than that of $L_{\mathrm{R}}$
(equal to \SI{300}{\pico\henry}). Accordingly, the reality of $\Omega_{\mathrm{m}}\ll\omega_{\downarrow},\omega_{\uparrow}$
can also be found. A detailed discussion is provided in Appendix C. 

For a CRLH TL, an equivalent relative permeability related to the
material can be expressed in the form 

\begin{equation}
\mu=\frac{1}{\mu_{0}}\left(L_{\mathrm{R}}^{\prime}-\frac{1}{\omega^{2}C_{\mathrm{L}}^{\prime}}\right),\label{eqMu.material}
\end{equation}
where $\mu_{0}$ is the vacuum permeability, $L_{\mathrm{R}}^{\prime}$
is the per-unit length shunt inductance, and $C_{\mathrm{L}}^{\prime}$
is the times-unit length series capacitance of the CRLH TL \citep{Lai2004IMM}.
Equation (\ref{eqMu.material}) tells that $\mu$ can be tuned if
the related $L_{\mathrm{R}}^{\prime}$ or $C_{\mathrm{L}}^{\prime}$
can be adjusted. Furthermore, the modulation of permeability in the
form of (\ref{eqModuMuDis}) is possible by the proper control of
$L_{\mathrm{R}}^{\prime}$ or $C_{\mathrm{L}}^{\prime}$. In this
work, we tune the mutual inductance $L_{\mathrm{m}}$ to realize the
modulation. The reason is that $L_{\mathrm{m}}$ equivalently contributes
to $L_{\mathrm{R}}$ for the even mode in the coupled-line {[}see
\figref{fig2}{b}{]}. Alternatively, from the perspective of coupling
coefficient, $K$ is modulated by varying $\beta_{\mathrm{even}}$
according to equation (\ref{eqCoupRate.even-odd}), which is essentially
implemented by the modulation of $L_{\mathrm{m}}$.

\section*{Appendix B: Basic analyses of periodic CRLH TLs}

\noindent 
\begin{figure*}
\centering\vspace{4mm}

\includegraphics{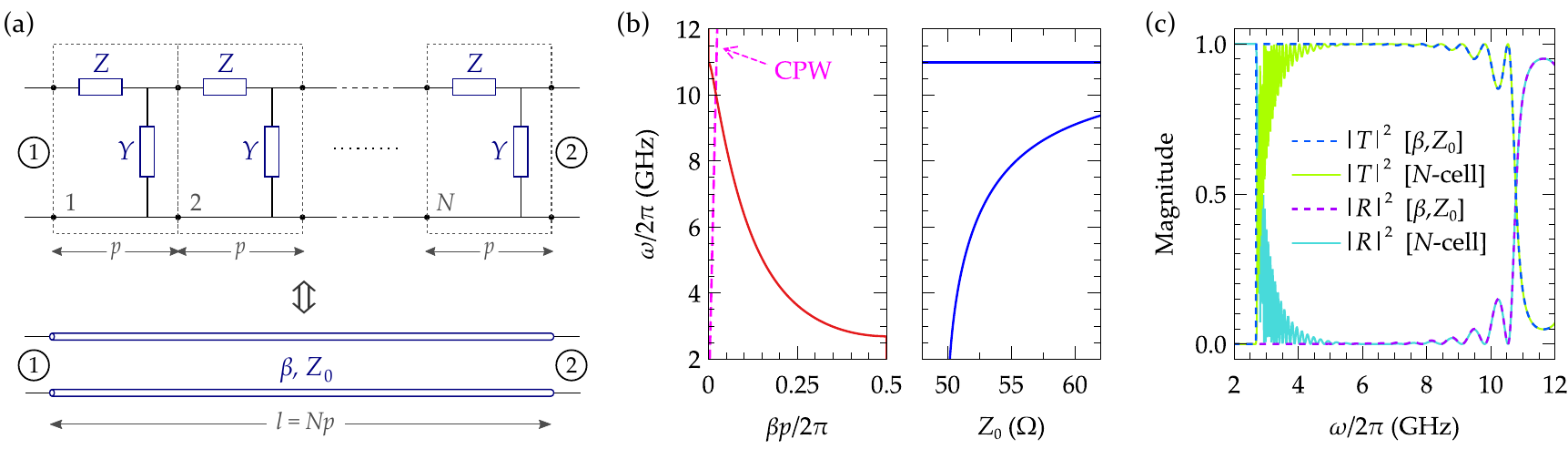} \caption{CRLH transmission line. (a) Periodic $N$-cell network and the equivalent
homogeneous TL in the long-wavelength limit. (b) Frequency dependence
of propagation constant (left) and characteristic impedance (right).
(c) Transmissions and reflections of CRLH TL with length of $40p$
and terminal impedances of \SI{50}{\ohm}. The calculations are performed
by two methods of $N$-cell network and homogeneous TL. \label{fig10}}
\end{figure*}

Here, we give the basic analyses of a periodic CRLH TL as shown in
\figref{fig10}{a} {[}see \figref{fig2}{a} for a unit cell{]}
\citep{Lai2004IMM,Martin2015Book}. By applying periodic boundary
conditions (PBCs) related to the Bloch-Floquet theorem to the unit
cell with length $p$, we can derive the CRLH TL propagation constant
as

\begin{equation}
\beta(\omega)=\frac{1}{p}\cos^{-1}\left[1+\frac{Z(\omega)Y(\omega)}{2}\right],\label{eqBeta.CRLH-TL}
\end{equation}
where the series impedance ($Z$) and shunt admittance ($Y$) of the
CRLH TL unit cell are respectively expressed in the forms 
\begin{align}
Z(\omega) & =\mathrm{j}\left(\omega L_{\mathrm{R}}-\frac{1}{\omega C_{\mathrm{L}}}\right),\\
Y(\omega) & =\mathrm{j}\left(\omega C_{\mathrm{R}}\text{\textminus}\frac{1}{\omega L\mathrm{_{L}}}\right).
\end{align}
Then, the characteristic impedance of the CRLH TL is given by 
\begin{equation}
Z_{0}(\omega)=\sqrt{Z(\omega)/Y(\omega)}.
\end{equation}
Figure \fref{fig10}{b} displays the dispersion curve and characteristic
impedance of a CRLH TL with the same $\{L_{\mathrm{R}},C_{\mathrm{R}}\}$
and $\{L_{\mathrm{L}},C_{\mathrm{L}}\}$ used in the main text. There
are two important features. First, the propagation constant is about
one order of magnitude greater than that of a common coplanar waveguide
at around the frequency of \SI{6}{\GHz}. This means that the effective
wavelength is roughly decreased by one order of magnitude, which implies
that the structure size can be reduced significantly with CRLH TLs.
Second, the characteristic impedance roughly equals \SI{50}{\ohm}
at around \SI{6}{\GHz}, which means that the impedance matching is
not a critical problem with the CRLH TLs.

To calculate the transmissions of CRLH TLs, the $ABCD$-matrix of
a symmetric unit cell is introduced and presented as follows:

\begin{equation}
\left[\begin{array}{cc}
A_{\mathrm{u}} & B_{\mathrm{u}}\\
C_{\mathrm{u}} & D_{\mathrm{u}}
\end{array}\right]=\left[\begin{array}{cc}
1+ZY/2 & Z+Z^{2}Y/4\\
Y & 1+ZY/2
\end{array}\right].
\end{equation}
For $N$ cells of the periodic CRLH TL, we have

\begin{equation}
\left[\begin{array}{cc}
A_{\mathrm{u}N} & B_{\mathrm{u}N}\\
C_{\mathrm{u}N} & D_{\mathrm{u}N}
\end{array}\right]=\left[\begin{array}{cc}
A_{\mathrm{u}} & B_{\mathrm{u}}\\
C_{\mathrm{u}} & D_{\mathrm{u}}
\end{array}\right]^{N}.\label{eqABCD.Ncells}
\end{equation}
Specifically, in the long-wavelength limit (i.e., $\beta p\ll1$),
Eq.\,(\ref{eqBeta.CRLH-TL}) is reduced to

\begin{equation}
\beta(\omega)=-\mathrm{j}\sqrt{Z(\omega)Y(\omega)},\label{eqBeta.HomTLs}
\end{equation}
which is known as the general expression of $\beta$ for a homogeneous
TL shown in \figref{fig10}{a}. Meanwhile, the $ABCD$-matrix of
a homogeneous TL with length $l$ is given by

\begin{equation}
\left[\begin{array}{cc}
A_{\mathrm{TL}} & B_{\mathrm{TL}}\\
C_{\mathrm{TL}} & D_{\mathrm{TL}}
\end{array}\right]=\left[\begin{array}{cc}
\cos(\beta l) & \mathrm{j}Z_{0}\sin(\beta l)\\
\mathrm{j}Z_{0}^{-1}\sin(\beta l) & \cos(\beta l)
\end{array}\right].\label{eqABCD.HomTLs}
\end{equation}
Once the $ABCD$-matrix for an $N$-cell network TL or homogeneous
TL has been established, the corresponding scattering parameters $S_{ij}$
for terminations of impedance $Z\mathrm{_{c}}$ can be computed by
using the well-known formula

\begin{widetext}

\begin{equation}
\begin{array}{c}
\left[\begin{array}{cc}
S_{11} & S_{12}\\
S_{21} & S_{22}
\end{array}\right]={\displaystyle \frac{1}{A+B/Z\mathrm{_{c}}+CZ\mathrm{_{c}}+D}}\left[\begin{array}{cc}
A+B/Z\mathrm{_{c}}-CZ\mathrm{_{c}}-D & 2\left(AD-BC\right)\\
2 & -A+B/Z\mathrm{_{c}}-CZ\mathrm{_{c}}+D
\end{array}\right]\end{array}.
\end{equation}

\end{widetext}

Here, for comparison, the transmission and reflection spectra are
calculated by using both the $N$-cell network expression in (\ref{eqABCD.Ncells})
and the homogeneous TL expression in (\ref{eqABCD.HomTLs}). The results
are shown in \figref{fig10}{c} and indicate that there is no significant
difference in the high-frequency region. Note that, from the dispersion
curve shown in \figref{fig10}{b}, the high-frequency region corresponds
to the long-wavelength region, where the long-wavelength approximation
is valid. From \figref{fig10}{c}, it is found that the spectra
obtained by two methods are matched well at around \SI{6}{\GHz}.
Thus, a good analysis and prediction can be given by the coupled-mode
theory for homogeneous TLs, then the accurate spectra can be obtained
by using the $N$-cell network calculations.

\section*{Appendix C: Asymmetrical Coupled-Line}

As described Appendix A, to realize the temporal modulation and efficient
transition between mode-$[\uparrow]$ and mode-{[}$\downarrow${]},
an asymmetric CRLH coupler is required in our design. The circuit
model of a unit cell of the symmetric CRLH coupler and its equivalent
even/odd line models are shown in \figref{fig2}{b} of the main
text. However, it is not easy to obtain an equivalent circuit for
an asymmetric CRLH coupler. Here, we analyze a homogeneous asymmetric
coupled-line on the basis of the coupled-mode theory. In an asymmetric
CRLH coupler, the normal modes become $c$ (even-like) and $\pi$
(odd-like) modes, which are not perfect even and odd modes \citep{Tripathi1975ITMTT,Islam2006IMWCL,Mmongia2007Book}.
The complex propagation constants $\gamma$ of the $c$ and $\pi$
modes are found to be

\begin{equation}
\begin{array}{l}
\left(\begin{array}{cc}
a-\gamma^{2} & b\\
c & d-\gamma^{2}
\end{array}\right)\left(\begin{array}{l}
V_{\uparrow}\\
V_{\downarrow}
\end{array}\right)=0\end{array}
\end{equation}
where

\[
\begin{array}{l}
a=Z_{\uparrow}Y_{\uparrow}+Z_{\mathrm{m}}Y_{\mathrm{m}},\quad b=Z_{\uparrow}Y_{\mathrm{m}}+Z_{\mathrm{m}}Y_{\downarrow}\\
c=Z_{\downarrow}Y_{\mathrm{m}}+Z_{\mathrm{m}}Y_{\uparrow},\quad d=Z_{\downarrow}Y_{\downarrow}+Z_{\mathrm{m}}Y_{\mathrm{m}}
\end{array}
\]
with series impedances $Z_{\uparrow,\downarrow}$ and shunt admittances
$Y_{\uparrow,\downarrow}$ of uncoupled line-$[\uparrow]$ and line-$[\uparrow]$,
$Z_{\mathrm{m}}=\mathrm{j}\omega L_{\mathrm{m}}$, and $Y_{\mathrm{m}}=-\mathrm{j}\omega C_{\mathrm{m}}$.
Considering a lossless coupled-line, the propagation constants (i.e.,
wavenumbers) of the $c$ and $\pi$ modes are solved in the form

\begin{align}
\beta_{c,\pi} & =\text{\textminus}\mathrm{j}\gamma_{c,\pi}\nonumber \\
 & =\text{\textminus}\mathrm{j}\left[\frac{a+d}{2}\pm\frac{1}{2}\sqrt{(a-d)^{2}+4bc}\right]^{1/2}.\label{eqBeta.CandP}
\end{align}

Specifically, for even and odd modes, this expression is reduced to
the identical form as Eq.\,(\ref{eqBeta.HomTLs}). Moreover, the
uncoupled eigenmodes of the individual lines are obtained by setting
$L_{\mathrm{m}}=C_{\mathrm{m}}=0$, and their isolated self-wavenumbers
$\beta_{\uparrow}=\text{\textminus}\mathrm{j}\sqrt{Z_{\uparrow}Y_{\uparrow}}$
and $\beta_{\downarrow}=\text{\textminus}\mathrm{j}\sqrt{Z_{\downarrow}Y_{\downarrow}}$.
Note that the self-wavenumbers for the individual lines in a coupled-line
is not exactly equal to that of the uncoupled line, since the self-inductance
and capacitance are modified in the presence of a neighboring line.
However, these modifications are minute and omitted, especially for
the consideration of the wavenumber difference $\Delta\beta_{0}=\beta_{\downarrow}-\beta_{\uparrow}$.
Here, we assume that the asymmetry of the two lines is introduced
by adding an $L_{\mathrm{R}}$-shift of $-6$ and \SI{+6}{\pico\henry}
for line-$[\uparrow]$ and line-{[}$\downarrow${]}, respectively.
The band structure of two uncoupled eigenmodes is just the plotting
diagram shown in Fig.\,\ref{fig9}. In the asymmetric coupler, the
$\Delta\beta_{\mathrm{c}}$ will be given by $\beta_{c}-\beta_{\pi}$,
instead of $\beta_{\mathrm{even}}-\beta_{\mathrm{odd}}$. With the
use of equation (\ref{eqBetasplitting.coupling}), the coupling coefficient
can be written as

\begin{equation}
K_{0}=\frac{\sqrt{\Delta\beta_{\mathrm{c}}^{2}-\Delta\beta_{0}^{2}}}{2}.
\end{equation}
Now, we deduce the responses of the entire structure with asymmetric
CRLH coupled-lines. Similarly to the treatment approach for the case
with symmetric couplers in the main text: first, we assume that there
is no modulation for the coupler (i.e., static) and obtain the required
$l_{\mathrm{c}}$ and scattering matrix for the 3-dB coupling at \SI{6}{\GHz}.
Second, we consider the modulated coupler, where a modulation initial
phase will be applied to the cross-coupling terms in the scattering
matrix of the static coupler. Finally, the total responses of the
entire structure can be obtained by a cascade multiplication of scattering
matrices of stage-I to stage-III. Calculations of the second and third
steps directly refer to the descriptions in Section \ref{secV} of
the main text. Here, we give the expressions of scattering parameters
of a static asymmetric coupled line that appeared in the first step.
Since there is no modulation, we can set $\Omega_{\mathrm{m}}=0$
and $\phi=0$ in Eq.\,(\ref{eqSolu_tsInter}) and deduce the corresponding
scattering parameters according to the definition of a scattering
matrix. The scattering parameters of the asymmetric static coupled
line are presented in the forms

\begin{widetext}

\begin{align}
M_{\uparrow\uparrow}^{(\mathrm{c})} & \equiv S_{\mathcal{A}\uparrow\rightarrow\mathcal{B}\uparrow}={\displaystyle \mathrm{e}^{-\mathrm{j}\frac{\Delta\beta_{\mathrm{0}}l_{\mathrm{c}}}{2}}{\displaystyle \left\{ \cos\left(\frac{\Delta\beta_{\mathrm{c}}l_{\mathrm{c}}}{2}\right)+\mathrm{j}\frac{\Delta\beta_{0}}{\Delta\beta_{\mathrm{c}}}\sin\left(\frac{\Delta\beta_{\mathrm{c}}l_{\mathrm{c}}}{2}\right)\right\} }},\label{eqMc11.static-1}\\
M_{\downarrow\uparrow}^{(\mathrm{c})} & \equiv S_{\mathcal{A}\uparrow\rightarrow\mathcal{B}\downarrow}=\mathrm{-j}\mathrm{e}^{-\mathrm{j}\frac{\Delta\beta_{\mathrm{0}}l_{\mathrm{c}}}{2}}{\displaystyle \frac{2K_{0}}{\Delta\beta_{\mathrm{c}}}\sin\left(\frac{\Delta\beta_{\mathrm{c}}l_{\mathrm{c}}}{2}\right)},\label{eqMc21.static-1}\\
M_{\uparrow\downarrow}^{(\mathrm{c})} & \equiv S_{\mathcal{A}\downarrow\rightarrow\mathcal{B}\uparrow}=\mathrm{-j}\mathrm{e}^{\mathrm{j}\frac{\Delta\beta_{\mathrm{0}}l_{\mathrm{c}}}{2}}{\displaystyle \frac{2K_{0}}{\Delta\beta_{\mathrm{c}}}\sin\left(\frac{\Delta\beta_{\mathrm{c}}l_{\mathrm{c}}}{2}\right)},\label{eqMc12.static-1}\\
M_{\downarrow\downarrow}^{(\mathrm{c})} & \equiv S_{\mathcal{A}\downarrow\rightarrow\mathcal{B}\downarrow}=\mathrm{e}^{\mathrm{j}\frac{\Delta\beta_{\mathrm{0}}l_{\mathrm{c}}}{2}}{\displaystyle \left\{ \cos\left(\frac{\Delta\beta_{\mathrm{c}}l_{\mathrm{c}}}{2}\right)-\mathrm{j}\frac{\Delta\beta_{0}}{\Delta\beta_{\mathrm{c}}}\sin\left(\frac{\Delta\beta_{\mathrm{c}}l_{\mathrm{c}}}{2}\right)\right\} }.\label{eqMc22.static-1}
\end{align}

\end{widetext}

\begin{figure}
\centering 

\includegraphics[width=8.5cm]{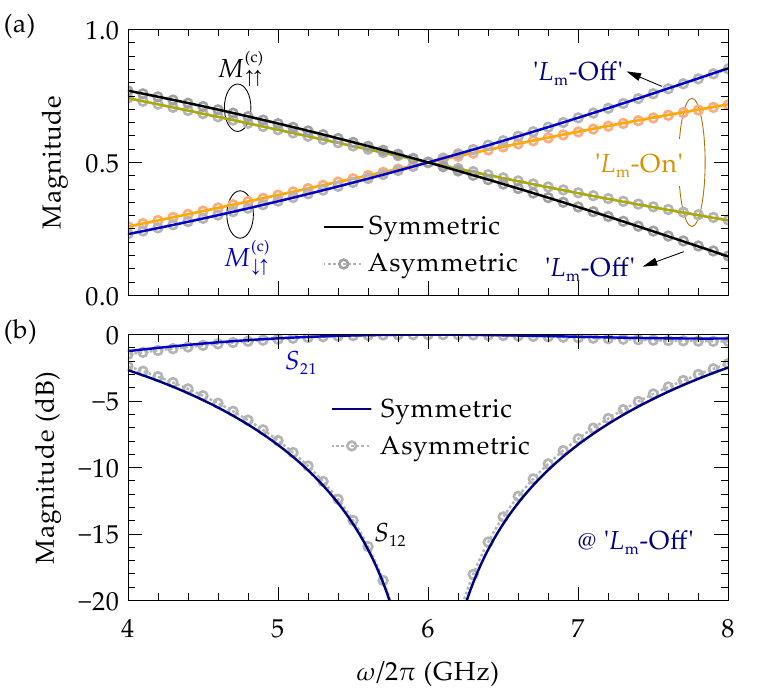} \caption{Responses using asymmetric and symmetric directional couplers. (a)
Responses of 3-dB coupler under the conditions of both $L_{\mathrm{m}}$-On
and $L_{\mathrm{m}}$-Off. (b) Nonreciprocal isolation response under
the condition of $L_{\mathrm{m}}$-Off. \label{fig11}}
\end{figure}

By using Eqs.\,(\ref{eqBeta.CandP}--\ref{eqMc22.static-1}), the
calculated scattering parameters are shown in \figref{fig11}{a}
and depict the realization of 3-dB coupling. By substituting these
results into the equations presented in Section \ref{secV}, we obtain
the results of total scattering parameters displayed in \figref{fig11}{b}.
As a comparison, we also demonstrate the results of the symmetric
coupled-line with the model of homogeneous TLs. It is found that these
differences are negligible, owing to the very small variation of the
coupling coefficient induced by a small $L_{\mathrm{R}}$-shift (i.e.,
$\Delta\beta_{0}\ll K_{0}$). It should be noted that even an asymmetric
coupled-line is used in the real design, but the interband transition
effectively occurs at $\Delta\beta_{0}\approx0$ under a dynamic modulation.
The demonstration here using the scattering matrix of a static coupler
with $\Omega_{\mathrm{m}}$ \emph{equal to 0} in the first step, provides
an estimation (or a perspective) of the effect induced by a small
asymmetry in the structure, but this is not the real process in the
interband transition with $\Omega_{\mathrm{m}}\neq0$.

\section*{Appendix D: Scattering Matrix of Interconnected Networks}

\begin{figure}
\centering

\includegraphics{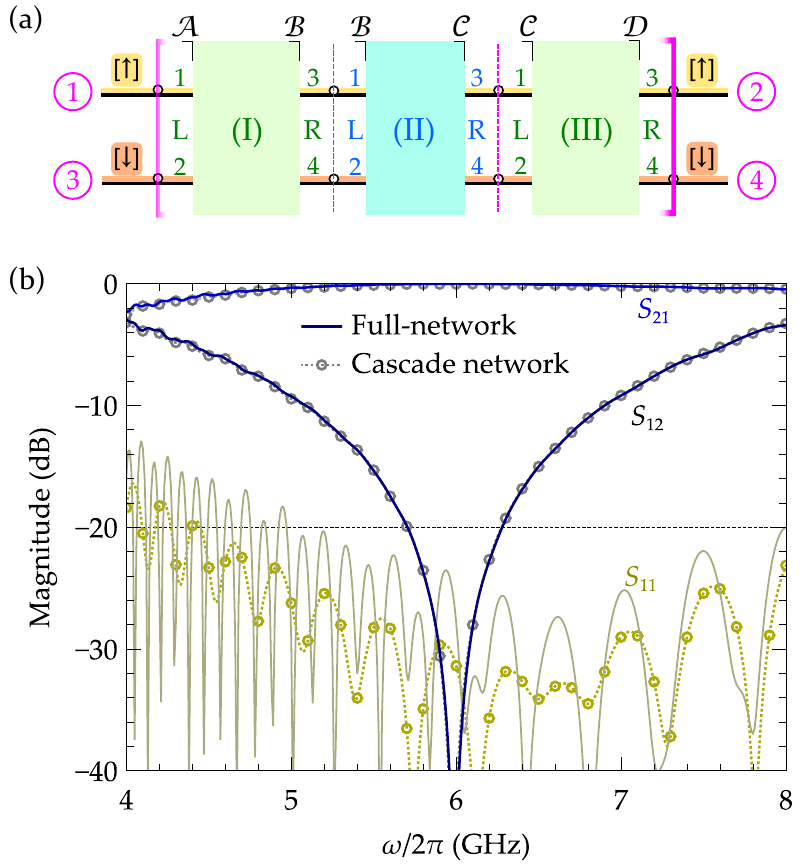} \caption{Results of full-network (four-port) and cascade network (two-port).
(a) Schematic for full-network analysis. (b) Results of nonreciprocal
isolation calculated by full-network and cascade network methods.
\label{fig12}}
\end{figure}

In the main text, we calculate the total responses with cascade multiplications
of scattering matrices of stage-I, stage-II, and stage-III. For each
stage, the scattering matrix is expressed by a $2\times2$ matrix
used for the two-port network. With such an expression, we have assumed
that there are no reflections within TLs, so a physical four-port
structure can be reduced to a two-port network. However, some weak
inter-reflections may be induced by connections between neighboring
stages, which are not considered in the cascade two-port network.
To obtain an accurate response for the real full (four-port) network,
we give a calculation approach with interconnected networks \citep{Mmongia2007Book}.
First, for each stage, we separate the four ports into the left-subgroup
(including port-1 and port-2) and the right-subgroup (including port-3
and port-4), as shown in \figref{fig12}{a}. Accordingly, the scattering
matrix of stage-I can be expressed in the form

\begin{widetext}

\begin{equation}
\mathbf{S}^{(\mathrm{I})}\equiv\left[\begin{array}{cc}
[\mathbf{S}_{\mathrm{LL}}^{(\mathrm{I})}] & [\mathbf{S}_{\mathrm{LR}}^{(\mathrm{I})}]\\{}
[\mathbf{S}_{\mathrm{RL}}^{(\mathrm{I})}] & [\mathbf{S}_{\mathrm{RR}}^{(\mathrm{I})}]
\end{array}\right]=\left[\begin{array}{cc|cc}
\varGamma & \varUpsilon & M_{\uparrow\uparrow}^{(\mathrm{c})} & \mathrm{e}^{-\mathrm{j}\phi_{1}}M_{\uparrow\downarrow}^{(\mathrm{c})}\\
\varUpsilon & \varGamma & \mathrm{e}^{\mathrm{j}\phi_{1}}M_{\downarrow\uparrow}^{(\mathrm{c})} & M_{\downarrow\downarrow}^{(\mathrm{c})}\\
\hline M_{\uparrow\uparrow}^{(\mathrm{c})} & \mathrm{e}^{-\mathrm{j}\phi_{1}}M_{\uparrow\downarrow}^{(\mathrm{c})} & \varGamma & \varUpsilon\\
\mathrm{e}^{\mathrm{j}\phi_{1}}M_{\downarrow\uparrow}^{(\mathrm{c})} & M_{\downarrow\downarrow}^{(\mathrm{c})} & \varUpsilon & \varGamma
\end{array}\right],\label{eqScatM.4p-I}
\end{equation}
the scattering matrix for stage-II 

\begin{equation}
\mathbf{S}^{(\mathrm{II})}\equiv\left[\begin{array}{cc}
[\mathbf{S}_{\mathrm{LL}}^{(\mathrm{II})}] & [\mathbf{S}_{\mathrm{LR}}^{(\mathrm{II})}]\\{}
[\mathbf{S}_{\mathrm{RL}}^{(\mathrm{II})}] & [\mathbf{S}_{\mathrm{RR}}^{(\mathrm{II})}]
\end{array}\right]=\left[\begin{array}{cc|cc}
0 & 0 & S_{\mathcal{CB}}^{[\mathrm{\uparrow}]} & 0\\
0 & 0 & 0 & S_{\mathcal{CB}}^{[\mathrm{\downarrow}]}\\
\hline S_{\mathcal{CB}}^{[\mathrm{\uparrow}]} & 0 & 0 & 0\\
0 & S_{\mathcal{CB}}^{[\mathrm{\downarrow}]} & 0 & 0
\end{array}\right],\label{eqScatM.4p-II}
\end{equation}
and the scattering matrix for stage-III

\begin{equation}
\mathbf{S}^{(\mathrm{III})}\equiv\left[\begin{array}{cc}
[\mathbf{S}_{\mathrm{LL}}^{(\mathrm{III})}] & [\mathbf{S}_{\mathrm{LR}}^{(\mathrm{III})}]\\{}
[\mathbf{S}_{\mathrm{RL}}^{(\mathrm{III})}] & [\mathbf{S}_{\mathrm{RR}}^{(\mathrm{III})}]
\end{array}\right]=\left[\begin{array}{cc|cc}
\varGamma & \varUpsilon & M_{\uparrow\uparrow}^{(\mathrm{c})} & \mathrm{e}^{\mathrm{-j}\phi_{2}}M_{\uparrow\downarrow}^{(\mathrm{c})}\\
\varUpsilon & \varGamma & \mathrm{e}^{\mathrm{j}\phi_{2}}M_{\downarrow\uparrow}^{(\mathrm{c})} & M_{\downarrow\downarrow}^{(\mathrm{c})}\\
\hline M_{\uparrow\uparrow}^{(\mathrm{c})} & \mathrm{e}^{-\mathrm{j}\phi_{2}}M_{\uparrow\downarrow}^{(\mathrm{c})} & \varGamma & \varUpsilon\\
\mathrm{e}^{\mathrm{j}\phi_{2}}M_{\downarrow\uparrow}^{(\mathrm{c})} & M_{\downarrow\downarrow}^{(\mathrm{c})} & \varUpsilon & \varGamma
\end{array}\right].\label{eqScatM.4p-III}
\end{equation}

\end{widetext}

As the connection shown in \figref{fig12}{a}, for two neighboring
stages, the right-subgroup ports of the left-stage network are interconnected
to the left-subgroup ports of the right-stage network. For such an
interconnection, the scattering matrix for the connected network can
be deduced recursively by

\begin{equation}
\begin{aligned}\mathbf{S}_{\mathrm{LL}}^{(\mathrm{\imath\leftrightarrow\jmath})} & =\mathbf{S}_{\mathrm{LL}}^{(\imath)}+\mathbf{S}_{\mathrm{LR}}^{(\imath)}\left(\mathbf{I}-\mathbf{S}_{\mathrm{LL}}^{(\jmath)}\mathbf{S}_{\mathrm{RR}}^{(\imath)}\right)^{-1}\mathbf{S}_{\mathrm{LL}}^{(\jmath)}\mathbf{S}_{\mathrm{RL}}^{(\imath)},\\
\mathbf{S}_{\mathrm{LR}}^{(\mathrm{\imath\leftrightarrow\jmath})} & =\mathbf{S}_{\mathrm{LR}}^{(\imath)}\left(\mathbf{I}-\mathbf{S}_{\mathrm{LL}}^{(\jmath)}\mathbf{S}_{\mathrm{RR}}^{(\imath)}\right)^{-1}\mathbf{S}_{\mathrm{LR}}^{(\jmath)},\\
\mathbf{S}_{\mathrm{RL}}^{(\mathrm{\imath\leftrightarrow\jmath})} & =\mathbf{S}_{\mathrm{RL}}^{(\jmath)}\left(\mathbf{I}-\mathbf{S}_{\mathrm{RR}}^{(\imath)}\mathbf{S}_{\mathrm{LL}}^{(\jmath)}\right)^{-1}\mathbf{S}_{\mathrm{RL}}^{(\imath)},\\
\mathbf{S}_{\mathrm{RR}}^{(\mathrm{\imath\leftrightarrow\jmath})} & =\mathbf{S}_{\mathrm{RR}}^{(\jmath)}+\mathbf{S}_{\mathrm{RL}}^{(\jmath)}\left(\mathbf{I}-\mathbf{S}_{\mathrm{RR}}^{(\imath)}\mathbf{S}_{\mathrm{LL}}^{(\jmath)}\right)^{-1}\mathbf{S}_{\mathrm{RR}}^{(\imath)}\mathbf{S}_{\mathrm{LR}}^{(\jmath)},
\end{aligned}
\label{eqScatM.recur}
\end{equation}
where the superscripts $\imath$ and $\jmath$ are the indices of
connected left- and right-stage networks, respectively, $\mathbf{I}$
is a $2\times2$ identity matrix, and $()^{-1}$ denotes the inverse
of a matrix. By using these recurrence formulas of (\ref{eqScatM.recur}),
we can obtain the scattering matrix of the combined stages of I and
II (i.e., $\mathrm{I\leftrightarrow II}$) as 

\begin{equation}
\mathbf{S}^{(\mathrm{I\leftrightarrow II})}\equiv\left[\begin{array}{cc}
[\mathbf{S}_{\mathrm{LL}}^{(\mathrm{I\leftrightarrow II})}] & [\mathbf{S}_{\mathrm{LR}}^{(\mathrm{\mathrm{I\leftrightarrow II}})}]\\{}
[\mathbf{S}_{\mathrm{RL}}^{(\mathrm{I\leftrightarrow II})}] & [\mathbf{S}_{\mathrm{RR}}^{(\mathrm{I\leftrightarrow II})}]
\end{array}\right]\label{eqScatM.stageI-II}
\end{equation}
with settings $(\imath)=(\mathrm{I)}$ and $(\jmath)=\mathrm{(II)}$
in Eq.\,(\ref{eqScatM.recur}) and substitutions of Eqs.\,(\ref{eqScatM.4p-I})
and (\ref{eqScatM.4p-II}). Furthermore, the scattering matrix for
the entire structure (i.e., $\mathrm{I\leftrightarrow III}$) can
be achieved,

\begin{equation}
\mathbf{S}^{(\mathrm{I\leftrightarrow III})}\equiv\left[\begin{array}{cc}
[\mathbf{S}_{\mathrm{LL}}^{(\mathrm{I\leftrightarrow III})}] & [\mathbf{S}_{\mathrm{LR}}^{(\mathrm{\mathrm{I\leftrightarrow III}})}]\\{}
[\mathbf{S}_{\mathrm{RL}}^{(\mathrm{I\leftrightarrow III})}] & [\mathbf{S}_{\mathrm{RR}}^{(\mathrm{I\leftrightarrow III})}]
\end{array}\right]\label{eqScatM.stageI-III}
\end{equation}
with settings $(\imath)=(\mathrm{I\leftrightarrow II)}$ and $(\jmath)=\mathrm{(III)}$
in Eq.\,(\ref{eqScatM.recur}) and substitutions of Eqs.\,(\ref{eqScatM.stageI-II})
and (\ref{eqScatM.4p-III}). Finally, we rearrange the indices of
port-1 to port-4 for the entire structure as that in the main text
and provide a new notation for each scattering parameter (element)
in expression (\ref{eqScatM.stageI-III}) as

\begin{equation}
\mathbf{S}^{(\mathrm{I\leftrightarrow III})}\eqqcolon\left[\begin{array}{cccc}
S_{11}^{(\mathrm{tot)}} & S_{13}^{(\mathrm{tot)}} & S_{12}^{(\mathrm{tot)}} & S_{14}^{(\mathrm{tot)}}\\
S_{31}^{(\mathrm{tot)}} & S_{33}^{(\mathrm{tot)}} & S_{32}^{(\mathrm{tot)}} & S_{34}^{(\mathrm{tot)}}\\
S_{21}^{(\mathrm{tot)}} & S_{23}^{(\mathrm{tot)}} & S_{22}^{(\mathrm{tot)}} & S_{24}^{(\mathrm{tot)}}\\
S_{41}^{(\mathrm{tot)}} & S_{43}^{(\mathrm{tot)}} & S_{42}^{(\mathrm{tot)}} & S_{44}^{(\mathrm{tot)}}
\end{array}\right].\label{eqScatM.stageFull}
\end{equation}
With such a full-network (four-port) method, the accurate total scattering
parameters can be obtained. Figure \fref{fig12}{b} shows the calculated
results, and the results of the cascade network (two-port) method
are also demonstrated for comparison. It is found that the transmissions
obtained by two methods match well. The reflections have some difference
since the full network can reveal the inter-reflections between two
connections. It is found that the magnitude of reflection from the
full-network model is still less than \SI{-20}{\dB} within the frequency
region of interest (around \SI{6}{\GHz}).

%\bibliography{Bibfile}

%

\end{document}